\documentclass[a4paper, 11pt]{article}
\usepackage{amssymb,amsmath}
\usepackage{graphicx}
 \usepackage{latexsym,graphicx}
 \usepackage{float}
 \usepackage{amsmath}
 \usepackage{amsfonts}
 \usepackage{amssymb}
 \usepackage{epstopdf}
 \usepackage{color}
 \usepackage {hyperref}
 \usepackage{cite}
 \DeclareGraphicsExtensions{.eps}
 \numberwithin{equation}{section}
 
\begin{document}

 \newcommand{\be}[1]{\begin{equation}\label{#1}}
 \newcommand{\ee}{\end{equation}}
 \newcommand{\bea}{\begin{eqnarray}}
 \newcommand{\eea}{\end{eqnarray}}
 \def\disp{\displaystyle}

 \def\gsim{ \lower .75ex \hbox{$\sim$} \llap{\raise .27ex \hbox{$>$}} }
 \def\lsim{ \lower .75ex \hbox{$\sim$} \llap{\raise .27ex \hbox{$<$}} }

 \title{\Large \bf FLRW Cosmology in Metric-Affine F(R,Q) Gravity}
\author{Dinesh Chandra Maurya$^{1}$\footnote{Email: dcmaurya563@gmail.com (Corresponding Author)}, K. Yesmakhanova$^{2}$\footnote{Email: kryesmakhanova@gmail.com},   
R. Myrzakulov$^{2}$\footnote{Email: rmyrzakulov@gmai.com} , G. Nugmanova$^{2}$\footnote{Email: nugmanovagn@gmail.com}, \\ 
$^{1}$\textit{Centre for Cosmology, Astrophysics and Space Science,}\\ \textit{ GLA University, Mathura-281 406, Uttar Pradesh,India}\\
$^{2}$\textit{Ratbay Myrzakulov Eurasian International Centre for Theoretical} \\ \textit{Physics, Nur-Sultan, 010009, Kazakhstan}\\
}
\date{}
\maketitle
\date{}

 \begin{abstract}
      We investigate some FLRW cosmological models in the context of Metric-Affine $F(R,Q)$ gravity, as proposed in [arXiv:1205.52666]. Here, $R$ and $Q$ are the curvature and nonmetricity scalars using non-special connections, respectively. We get the modified field equations using a flat Friedmann-Lema\^{i}tre-Robertson-Walker (FLRW) metric. We then find a connection between the Hubble constant $H_{0}$, the density parameter $\Omega_{m0}$, and the other model parameters in two different situations involving scalars $u$ and $w$. Next, we used new observational datasets, such as the cosmic chronometer (CC) Hubble datasets and the Pantheon SNe Ia datasets, to determine the optimal model parameter values through MCMC analysis. Using these best-fit values of model parameters, we have discussed the results and behavior of the derived models. We have also discussed the AIC and BIC criteria for the derived models in the context of $\Lambda$CDM. We have found that the geometrical sector dark equation of state parameter $\omega_{de}$ behaves just like a dark energy candidate. We have found that both models are transit phase models and Model-I approaches to the Lambda CDM model in the late-time universe and Model-II approaches to quintessence scenarios.
\end{abstract}

\maketitle
	\smallskip
\vspace{5mm}
%\date{}
%\maketitle
\vspace{1cm}

\tableofcontents

%=================================================
\section{Introduction}
%=================================================

    Although General Relativity (GR) is unquestionably one of the most elegant and effective theories in physics, its position has been called into question by recent observational data \cite{ref1}. Perhaps the most significant observation is the fast expansion of our universe in early and late times \cite{ref2,ref3,ref4,ref5,ref6,ref7,ref8,ref9,ref10,ref11,ref12}, which defies explanation within the framework of general relativity. A variety of theories other than General Relativity (GR) have been developed as a result of this discrepancy between theory and observations; these theories are collectively referred to as Modified Gravity \cite{ref13}. We have demonstrated that the pursuit of a viable substitute has been beneficial and constructive for our understanding of gravity. There are many different types of modified gravities, such as metric $f(R)$ theories, the Metric-Affine (Palatini) $f(R)$ gravity \cite{ref14,ref15,ref16}, the teleparallel $f(T)$ gravities \cite{ref17,ref18}, the symmetric teleparallel $f(Q)$ \cite{ref19,ref20,ref21,ref22}, the scalar-tensor theories \cite{ref23,ref24}, and many more. Naturally, one's choice of alterations is very much a question of personal preference. From our perspective, intriguing and highly motivated alternatives are those that provide a more general connection than the typical Levi-Civita one, thus extending the fundamental geometry of spacetime. If there are no a priori limits on the connection, the space will usually not be Riemannian \cite{ref25} and will have both torsion and non-metricity. We conceptualize it as an additional fundamental field overlaying the metric. Identifying the affine relationship allows for the calculation of the final geometric quantities. Metric-Affine gravity theories are developed on this non-Riemannian manifold \cite{ref26}. Recently, \cite{ref27} talked about $f(R)$ gravity theories with a symmetric connection that is torsion-free. These are also known as Palatini $f(R)$ theories of gravity. U4 theories are $f(R)$ theories of gravity with torsion but don't have non-metricity. The dynamics of metric-affine gravity theories are discussed in \cite{ref28}, and the dynamics of generalized Palatini gravity theories are studied in \cite{ref29}. In \cite{ref30}, metric-affine variational principles in general relativity are talked about. Most recently, in \cite{ref31}, the role of non-metricity in metric-affine theories of gravity was studied into \cite{ref32}.\\
	
	The Metric-Affine technique \cite{ref16,ref17,ref18,ref19,ref20,ref21,ref22,ref23,ref24,ref25,ref26,ref27, ref28,ref29,ref30,ref31,ref32,ref33,ref34,ref35,ref36,ref37,ref38,ref39,ref40,ref41,ref42,ref43,ref44,ref45,ref46,ref47} has received significant attention in recent years, particularly for its cosmological applications \cite{ref48,ref49,ref50,ref51,ref52,ref53,ref54,ref55,ref56,ref57,ref58,ref59}. This interest may stem from the straightforward geometrical interpretation of the additional impacts that operate in this framework (in comparison to GR). In other words, spacetime torsion and non-metricity alone are responsible for the alterations. Moreover, matter with inherent structure excites these geometric concepts \cite{ref50} and \cite{ref60,ref61,ref62,ref63}. The MAG scheme gains an additional favorable aspect from this relationship between generalized geometry and inner structure. \\
	
	These, in turn, give us a reason to develop cosmological models in these affinely connected metric theories, especially from their Riemann-Cartan subclass \cite{ref64}, using a certain but not unique connection. By creating both non-zero curvature and non-zero torsion at the same time, this would add the extra degrees of freedom that are usually needed for any gravitational change \cite{ref65}. As a result, both early and late universe evolution may be satisfactorily explained by Metric-Affine gravity \cite{ref66,ref67,ref68,ref69,ref70}. \cite{ref66} is a new study of cosmology that was made possible by using this kind of framework and computing how observable quantities like density parameters and the effective dark energy equation-of-state parameter change over time. They have explored the cosmological behavior, emphasizing the connection's effect, using the mini-super-space technique, and expressing the theory as a deformation from both general relativity and its teleparallel counterpart. The observational limitations on Metric-Affine $F(R,T)$-gravity have been studied by \cite{ref71}. Several Metric-Affine Gravity Theories and their applications are discussed in \cite{ref72,ref73,ref74,ref75,ref76,ref77,ref78}.\\
	
	Motivated by the above discussions, we develop some FLRW cosmological models in torsion-free metric-affine geometry. We recently investigated transit phase cosmological models in Metric-Affine $F(R,T)$ gravity with observational constraints \cite{ref79}. In \cite{ref80,ref81}, we looked into some exact cosmological models in this metric-affine $F(R,T)$ gravity. In this paper, we investigate some FLRW cosmological models and their properties in the Metric-Affine $F(R,Q)$ gravity theory. For this, we consider the arbitrary function $F(R,Q)=R+\lambda Q+\lambda_{0}$, where $R$ is the Ricci scalar curvature, $Q$ is the non-metricity scalar with respect to non-special connection, and $\lambda, \lambda_{0}$ are arbitrary constants. \\
	
	The organization of the present paper is as follows: Sect.-2 presents some geometrical concepts of metric-affine spacetime, while Sect.-3 provides a brief introduction to the Metric-Affine $F(R,Q)$ gravity. We got the gravity field equations from the $F(R,Q)$ gravity theory and used them to study cosmological field equations of $F(R,Q)$ gravity in a flat FLRW spacetime in Sect.-4. In Sect. 5, we obtained two exact solutions of the derived field equations for different choices of $u$ and $w$. We have made observational constraints on the models obtained using two recent datasets $H(z)$ and Pantheon SNe Ia datasets by applying MCMC analysis in Sect.-6. Sect. 7 explores the results, while Sect. 8 presents the conclusions.

%=========================================================
\section{Geometrical preliminaries}
%=========================================================
The notion of metric-affine gravity is a generalization of the underlying connection. In this work, we generalize the connection in such a way that the torsion tensor $T_{\,\,\,\, \mu \nu}^{\alpha}$ should vanish (Weyl-type geometry). Therefore, such a connection can be defined as \cite{ref31}
\begin{equation}\label{2.1}
	\Gamma^{\rho}_{\,\,\,\,\mu\nu}=\breve{\Gamma}_{\,\,\,\, \mu \nu}^{\rho}+L^{\rho}_{\,\,\,\,\mu\nu},
\end{equation}
where $\Gamma^{\rho}_{\,\,\,\,\mu\nu}$ is called as symmetric general affine connection, $\breve{\Gamma}_{\,\,\,\, \mu \nu}^{\rho}$ is the Levi--Civita connection and $L^{\rho}_{\,\,\,\,\mu\nu}$ is  the disformation tensor. These two tensors have the following forms
\begin{eqnarray}
	\breve{\Gamma}^l_{\, \, \, jk} &=& \tfrac{1}{2} g^{lr} \left( \partial _k g_{rj} + \partial _j g_{rk} - \partial _r g_{jk} \right), \label{2.2} \\
	L^{\rho}_{\,\,\,\,\mu\nu}&=&\frac{1}{2}g^{\rho\lambda}\bigl(-Q_{\mu \nu \lambda}-Q_{\nu \mu \lambda} + Q_{\lambda\mu\nu}\bigr)=
	L^{\rho}_{\,\,\,\,\nu\mu}. \label{2.3}
\end{eqnarray}
where $Q_{\rho \mu \nu} = \nabla_{\rho} g_{\mu \nu}$ is the nonmetricity tensor.\\
Hence, we can expressed the Ricci curvature tensor $R_{\mu\nu}$ in terms of symmetric metric-affine connection \cite{ref74,ref82,ref83}, as below
\begin{equation}
	R_{\mu\nu}=\partial_{\lambda}\Gamma^{\lambda}_{\mu\nu}-\partial_{\mu}\Gamma^{\lambda}_{\lambda\nu}+\Gamma^{\lambda}_{\lambda\alpha}\Gamma^{\alpha}_{\mu\nu}-\Gamma^{\lambda}_{\mu\alpha}\Gamma^{\alpha}_{\lambda\nu}, \label{2.4}
\end{equation}
or
\begin{equation}
	R_{\mu\nu}=\breve{R}_{\mu\nu}+\partial_{\lambda}L^{\lambda}_{\mu\nu}-\partial_{\mu}L^{\lambda}_{\lambda\nu}+\breve{\Gamma}^{\lambda}_{\lambda\alpha}L^{\alpha}_{\mu\nu}+\breve{\Gamma}^{\alpha}_{\mu\nu}L^{\lambda}_{\lambda\alpha}-\breve{\Gamma}^{\lambda}_{\mu\alpha}L^{\alpha}_{\lambda\nu}-\breve{\Gamma}^{\alpha}_{\lambda\nu}L^{\lambda}_{\mu\alpha}+L^{\lambda}_{\lambda\alpha}L^{\alpha}_{\mu\nu}-L^{\lambda}_{\mu\alpha}L^{\alpha}_{\lambda\nu}, \label{2.5}
\end{equation}
where $\breve{R}_{\mu\nu}$ is the Ricci curvature tensor with respect to Levi-Civita connection $\breve{\Gamma}$. Now, the Ricci scalar $R$ with respect to general symmetric metric-affine connection $\Gamma$ can be expressed as
\begin{equation}
	R=\breve{R}+u, \label{2.6}
\end{equation}
where $u=u(\Gamma^{\rho}_{\,\,\,\,\mu\nu}, x_i, g_{ij}, \dot{g_{ij}},\ddot{g_{ij}}, ... , f_j)$ is a real function.\\
Similarly, we can expressed the nonmetricity tensor $Q_{\rho\mu\nu}$ with respect to general symmetric metric connection $\Gamma$ and in the case of non-coincident gauge formulation (see \cite{ref82,ref83,ref84}) as
\begin{equation}
	Q_{\rho\mu\nu}=\partial_{\rho}g_{\mu\nu}-\Gamma^{\lambda}_{\mu\rho}g_{\lambda\nu}-\Gamma^{\lambda}_{\nu\rho}g_{\lambda\mu}, \label{2.7}
\end{equation}
or
\begin{equation}
	Q_{\rho\mu\nu}=\breve{Q}_{\rho\mu\nu}+(-L^{\lambda}_{\mu\rho}g_{\lambda\nu}-L^{\lambda}_{\nu\rho}g_{\lambda\mu}). \label{2.8}
\end{equation}
Hence, the nonmetricity scalar $Q$ can be expressed as
\begin{equation}
	Q=\breve{Q}+w, \label{2.9}
\end{equation}
where $w=w(\Gamma^{\rho}_{\,\,\,\,\mu\nu}, x_i,  g_{ij}, \dot{g_{ij}},\ddot{g_{ij}}, ... , h_j)$ is a real function.\\
We will now introduce two geometrical scalars.
\begin{eqnarray}
 R&=&g^{\mu\nu}R_{\mu\nu},\label{2.10}\\
Q&=& -g^{\mu\nu}(L^{\alpha}_{\beta\mu}L^{\beta}_{\nu\alpha}-L^{\alpha}_{\beta\alpha}L^{\beta}_{\mu\nu}), \label{2.11}
 \end{eqnarray}  
where $R$ is the curvature scalar and $Q$ is the nonmetricity scalar. Here, $u$ may be a function of $w$.

%=========================================================
\section{Metric-Affine $F(R,Q)$ gravity}
%=========================================================

In the present work, we consider the Metric-Affine $F(R,Q)$ gravity \cite{ref85}. In this paper, we use the definitions and notations of \cite{ref86}, so we go through the basic setup rather briefly here and refer the reader to \cite{ref86} for more details. The action for $F(R,Q)$-gravity is described in \cite{ref85} as:
 \begin{equation}\label{3.1}
 S=\frac{1}{2\kappa}\int \left[F(R, Q)+2\kappa L_m\right]\sqrt{-g}~d^4x,
 \end{equation}
where $F(R, Q)$ is an arbitrary function of the Ricci $R$ scalar and the nonmetricity scalar $Q$, $g$ is the determinant of $g_{\mu\nu}$, and $\mathcal{L}_m$ is the matter Lagrangian density.\\
It is an extension of both the $F(R)$ and $F(Q)$ theories. Indeed, the function $F=F(R,Q)$ is a generic function of the scalar curvature $R$ (of the general affine connection $\Gamma$) and of $Q$, where $Q$ is the non-metricity scalar. The two independent traces of $Q_{\alpha\mu\nu}$ are
 \begin{equation}\label{3.2}
 Q_{\alpha}=Q_{\alpha }{}^{\mu }{}_{\mu }\,,\quad \tilde{Q}_{\alpha }=Q^{\mu
 }{}_{\alpha \mu }.
 \end{equation}
The invariant non-metricity scalar is defined as a contraction of $Q_{\alpha\mu \nu }$ given by
 \begin{equation}\label{3.3}
 Q=-Q_{\alpha \mu \nu }P^{\alpha \mu \nu},
 \end{equation}
 where $P^{\alpha \mu \nu}$ is the non-metricity conjugate and given by
 \begin{eqnarray}\label{3.4}
 4P^{\alpha }{}_{\mu \nu } &=& -Q^{\alpha }{}_{\mu \nu } + 2Q_{(\mu \phantom{\alpha}\nu )}^{\phantom{\mu}\alpha } - Q^{\alpha }g_{\mu \nu } - \tilde{Q}^{\alpha }g_{\mu \nu }-\delta _{(\mu }^{\alpha }Q_{\nu )}\,.
 \label{super}
 \end{eqnarray}
 The metric field equations of the theory read as follows:
\begin{equation}\label{3.5}
- \frac{1}{2} g_{\mu \nu} F + F_R R_{(\mu \nu)} + F_Q L_{(\mu \nu)} + \hat{\nabla}_\lambda \left(F_Q {J^\lambda}_{(\mu \nu)} \right) + g_{\mu \nu} \hat{\nabla}_\lambda \left(F_Q \zeta^\lambda \right) = \kappa T_{\mu \nu} \,,
\end{equation}
 where $F_{R}=\frac{\partial F}{\partial R}$, $F_{Q}=\frac{\partial F}{\partial Q}$ and $T_{\mu \nu }=-\frac{2}{\sqrt{-g}}\frac{\delta \left( \sqrt{-g}\mathcal{L}_{m}\right) }{\delta g^{\mu \nu }}$,  
\begin{equation}\label{3.6}
\hat{\nabla}_\lambda := \frac{1}{\sqrt{-g}}(2S_{\lambda}-\nabla_\lambda)
\end{equation}
and 
\begin{equation}\label{3.7}
\begin{split}
L_{\mu \nu} & := \frac{1}{4} \left[ \left(Q_{\mu \alpha \beta} - 2 Q_{\alpha \beta \mu} \right) {Q_\nu}^{\alpha \beta} + \left( Q_\mu + 2 \tilde{Q}_\mu \right) Q_\nu + \left( 2 Q_{\mu \nu \alpha} - Q_{\alpha \mu \nu} \right) Q^\alpha \right] \\
& \phantom{:= \,} - {\Xi^{\alpha \beta}}_\nu Q_{\alpha \beta \mu} - \Xi_{\alpha \mu \beta} {Q^{\alpha \beta}}_\nu \,, \\
{J^\lambda}_{\mu \nu} & := \sqrt{-g} \left(\frac{1}{4} {Q^\lambda}_{\mu \nu} - \frac{1}{2} {Q_{\mu \nu}}^\lambda + {\Xi^\lambda}_{\mu \nu} \right) \,, \\
\zeta^\lambda & := \sqrt{-g} \left( - \frac{1}{4} Q^\lambda + \frac{1}{2} \tilde{Q}^\lambda \right) \,, 
\end{split}
\end{equation}
where $Q_{\lambda \mu \nu}$ is the non-metricity tensor, $Q_\lambda$ and $\tilde{Q}_\lambda$ are its trace parts, and $\Xi_{\lambda \mu \nu}$ is the so-called (non-metricity) ``superpotential". The connection field equations are
\begin{equation}\label{3.8}
	{P_\lambda}^{\mu \nu} (F_R) + F_Q \left[ 2 {Q^{[\nu \mu]}}_\lambda - {Q_\lambda}^{\mu \nu} + \left( \tilde{Q}^\nu - Q^\nu \right) \delta^\mu_\lambda + Q_\lambda g^{\mu \nu} + \frac{1}{2} Q^\mu \delta^\nu_\lambda \right] = 0 \,,
\end{equation}
where ${P_\lambda}^{\mu \nu} (F_R)$ is  the modified Palatini tensor:
\begin{equation}\label{3.9}
	{P_\lambda}^{\mu \nu} (F_R) := - \frac{\nabla_\lambda \left(\sqrt{-g} F_R g^{\mu \nu} \right)}{\sqrt{-g}} + \frac{\nabla_\alpha \left( \sqrt{-g} F_R g^{\mu \alpha}\delta^\nu_\lambda \right)}{\sqrt{-g}} \,,
\end{equation}
being $\nabla$ the covariant derivative associated with the general affine connection $\Gamma$.\\
We assume that the matter is a perfect fluid whose energy-momentum tensor $T_{\mu \nu }$ is given by 
 \begin{equation}\label{3.10}
 T_{\mu \nu }=(\rho +p)u_{\mu }u_{\nu }+pg_{\mu \nu }\,,
 \end{equation}
 where $u_{\mu }$ is the four-velocity satisfying the normalization condition
 $u_{\mu }u^{\mu }=-1$, $\rho $ and $p$ are the energy density and pressure
 of a perfect fluid respectively.\\
 
%============================================================================
\section{FLRW cosmological field equations of $F(R,Q)$ gravity}
%============================================================================
 
First, let us rewrite the action \eqref{3.1} as 
\begin{eqnarray}\label{4.1}
S=\frac{1}{2\kappa^{2}}\int \sqrt{-g}d^{4}x[F(R,Q)-\lambda_{1}(R-R_{s}-u)-\lambda_{3}(Q-Q_{s}-w)+2\kappa^{2}L_{m}].
\end{eqnarray}
The variations of the action with respect to $R, Q$ give $\lambda_{1} = F_{R}, \lambda_{3}=F_{Q}$, respectively. Thus, the action \eqref{4.1} takes the form
\begin{eqnarray}\label{4.2}
S=\frac{1}{2\kappa^{2}}\int \sqrt{-g}d^{4}x[F-F_{R}(R-R_{s}-u)-F_{Q}(Q-Q_{s}-w)+2\kappa^{2}L_{m}],
\end{eqnarray}
where we know from the Eqs. \eqref{2.6} and \eqref{2.9} that $u=u(g_{ij}, \dot{g}_{ij}, \ddot{g}_{ij}, ...),  \quad w=w(g_{ij}, \dot{g}_{ij}, \ddot{g}_{ij}, ...)$. We now consider the FLRW spacetime case with the metric 
\begin{equation}\label{4.3}
 ds^2=-N^{2} (t)dt^2+a^2(t)(dx^2+dy^2+dz^2),
 \end{equation}
where $a=a(t)$ represents the scale factor, $N(t)$ represents the lapse function, and $N(t)=1$ is assumed. Then integrating by parts gives the following action with the point-like FLRW Lagrangian 
\begin{eqnarray}\label{4.4}
S=\frac{1}{2\kappa^{2}}\int {\cal L}dt,
\end{eqnarray}
The point-like Lagrangian has the form
\begin{eqnarray}\label{4.5}
{\cal L}= a^{3} [F-F_{R}(R-R_{s}-u)-F_{Q}(Q-Q_{s}-w)+2\kappa^{2}L_{m}].
\end{eqnarray}
In FLRW spacetime, we have
\begin{eqnarray}
R_{s}&=&6(\frac{\ddot{a}}{a}+\frac{\dot{a}^{2}}{a^{2}})=6(2H^{2}+\dot{H}), \label{4.6}\\
Q_{s}&=&6\frac{\dot{a}^{2}}{a^{2}}=6H^{2}. \label{4.7}
\end{eqnarray}
Finally, we get the following: FLRW Lagrangian
\begin{eqnarray}\label{4.8}
{\cal L}(a,R,Q,\dot{a},\dot{R},\dot{Q})= a^{3}(F-RF_{R}-QF_{Q})+6a\dot{a}^{2}(F_{R}+F_{Q})+6a^{2}\dot{a}\dot{F}_{R}+a^{3}(uF_{R}+wF_{Q})+2\kappa^{2}a^{3}L_{m},
\end{eqnarray}
Now, take the Hamiltonian ${\cal H}$ of the Lagrangian ${\cal L}$ as 
\begin{eqnarray}\label{4.9}
	{\cal H}={\cal E}=\dot{a}\frac{\partial {\cal L}}{\partial \dot{a}}+\dot{R}\frac{\partial {\cal L}}{\partial \dot{R}}+\dot{Q}\frac{\partial {\cal L}}{\partial \dot{Q}}-{\cal L}=0
\end{eqnarray}
and the Euler-Lagrange equations corresponding to the Lagrangian ${\cal L}$, we obtain the following field equations:
\begin{equation}\label{4.10}
	-\frac{1}{2}(F-RF_{R}-QF_{Q})+3H^{2}(F_{R}+F_{Q})-\frac{1}{2}[(u-\dot{a}u_{\dot{a}})F_{R}+(w-\dot{a}w_{\dot{a}})F_{Q}]+3H(\dot{R}F_{RR}+\dot{Q}F_{RQ})=\kappa^{2}\rho
\end{equation}
\begin{multline}\label{4.11}
	-\frac{1}{2}(F-RF_{R}-QF_{Q})+(2\dot{H}+3H^{2})(F_{R}+F_{Q})-\frac{1}{2}(u+\frac{1}{3}au_{a}-\dot{a}u_{\dot{a}}-\frac{1}{3}a\dot{u}_{\dot{a}})F_{R}-\frac{1}{2}(w+\frac{1}{3}aw_{a}-\dot{a}w_{\dot{a}}-\frac{1}{3}a\dot{w}_{\dot{a}})F_{Q}\\+2H(\dot{F}_{R}+\dot{F}_{Q})+\frac{1}{6}a(u_{\dot{a}}\dot{F}_{R}+w_{\dot{a}}\dot{F}_{Q})+\ddot{F}_{R}=-\kappa^{2}p,
\end{multline}
where
\begin{equation}\label{4.12}
	\rho=L_{m}-\dot{a}\frac{\partial L_{m}}{\partial \dot{a}},~~~~p=\frac{1}{3a^{2}}\left[\frac{d}{dt}\left(a^{3}\frac{\partial L_{m}}{\partial \dot{a}}\right)-\frac{\partial}{\partial a}(a^{3}L_{m})\right].
\end{equation}

 %=============================================================================
	\section{Cosmological solutions for $F(R,Q)=R+\lambda Q+\lambda_{0}$ gravity. }
 %=============================================================================

In the present work, we are interested in investigating the cosmological behavior that arises purely from the non-special symmetric connection of metric-affine gravity. We chose the simplest case, where the arbitrary function is simple: $F(R,Q)= R+\lambda Q+\lambda_{0}$, where $\lambda$ is a dimensionless parameter (we don't include the coupling coefficient of $R$ because it can be absorbed into $\kappa^{2}$) and $\lambda_{0}$ is an arbitrary constant (a model free parameter of dimension of $H_{0}^{2}$). As a result, we for this particular case of the arbitrary function $F(R,Q)= R+\lambda Q+\lambda_{0}$ with $\lambda, \lambda_{0}$ as model parameters, the field equations \eqref{4.10} \& \eqref{4.11} become
	\begin{equation}
		3(1+\lambda)H^{2}-\frac{1}{2}[(u-\dot{a}u_{\dot{a}})+\lambda(w-\dot{a}w_{\dot{a}})]-\frac{\lambda_{0}}{2}=\kappa^{2}\rho,  \label{5.1}
	\end{equation}
	\begin{equation}
		(1+\lambda)(2\dot{H}+3H^{2})-\frac{1}{2}[(u+\frac{1}{3}au_{a}-\dot{a}u_{\dot{a}}-\frac{1}{3}a\dot{u}_{\dot{a}})+\lambda(w+\frac{1}{3}aw_{a}-\dot{a}w_{\dot{a}}-\frac{1}{3}a\dot{w}_{\dot{a}})]-\frac{\lambda_{0}}{2}=-\kappa^{2}p,  \label{5.2}
	\end{equation}
At the same time, for the original density and pressure, the continuity equation takes the form
\begin{eqnarray}
 \dot{\rho}+3H(\rho+ p)+\frac{1}{2\kappa^{2}}(\dot{y}-\dot{a}y_{a}-\ddot{a}y_{\dot{a}})=0,  \label{5.3}
 \end{eqnarray}
where
\begin{eqnarray}
 y=u+\lambda w.  \label{5.4}
 \end{eqnarray}
 Now, we have two linearly independent field equations \eqref{5.1} and \eqref{5.2} in five unknowns: $\rho, p, a, u, w$. To find the exact solutions to these two field equations, we must impose at least three constraints on these unknowns. This modified $F(R,Q)$ gravity theory depends upon the choices of the factors $u(a,\dot{a}, \ddot{a},..)$ and $w(a,\dot{a}, \ddot{a},..)$, which can be considered as per their definitions (see \eqref{2.6} and \eqref{2.9}). Therefore, we investigate the above model using two different choices for $u$ and $w$, resulting in two distinct cosmological models, as detailed below.

%===================================================
\subsection{Model-I}
%===================================================

As we have discussed in Section 2, the scalars $u$ and $w$ may be functions of scale factor $a(t)$, connection $\Gamma$, and its derivatives. In our study, we choose the scalars $u$ and $w$ such that the energy conservation equation \eqref{5.3} is satisfied. Thus, to get the exact solutions to the field equations \eqref{5.1} and \eqref{5.2}, without loss of generality, we can pick $u=c_{1}\frac{\dot{a}}{a}\ln\dot{a}$ and $w=s(a)\dot{a}$, where $c_{1}$ is a constant and $s(a)$ is any function of $a$. There may be several such choices as per the concepts of $u, w$ (see \cite{ref71,ref79,ref80,ref81}). Then the above field equations \eqref{5.1} \& \eqref{5.2} become
\begin{equation}
	3(1+\lambda)H^{2}+\frac{1}{2}c_{1}H-\frac{1}{2}\lambda_{0}=\kappa^{2}\rho, \label{5.5}
\end{equation}
\begin{equation}
	(1+\lambda)(2\dot{H}+3H^{2})+\frac{1}{6}c_{1}\frac{\dot{H}}{H}+\frac{1}{2}c_{1}H-\frac{1}{2}\lambda_{0}=-\kappa^{2}p, \label{5.6}
\end{equation}
and the energy conservation equation \eqref{5.3} becomes
\begin{eqnarray}
	\dot{\rho}+3H(\rho+ p)=0,  \label{5.7}
\end{eqnarray}
Third constraint, we take on matter pressure as $p\approx0$, and solving the energy conservation equation \eqref{5.7}, we obtain the matter energy density $\rho$ as
\begin{equation}
	\rho=\rho_{0}\left(\frac{a_{0}}{a}\right)^{3}=\rho_{0}(1+z)^{3}, \label{5.8} 
\end{equation}
where $\rho_{0}$ is the present value of energy density $\rho$ at $z=0$ and $\frac{a_{0}}{a}=1+z$ with $a(t)$ as scale factor.\\
Now from \eqref{5.5}, we can find the relation at present ($z=0$) as
\begin{equation}
	6(1+\lambda)=6\Omega_{m0}+\frac{\lambda_{0}}{H_{0}^{2}}-\frac{c_{1}}{H_{0}}\label{5.9}
\end{equation}
where $\Omega_{m0}=\frac{\kappa^{2}\rho_{0}}{3H_{0}^{2}}$. This equation \eqref{5.9} suggests that $\lambda$ is a dimensionless parameter, $\lambda_{0}$ is a parameter of dimension $H_{0}^{2}$ and $c_{1}$ is a parameter of dimension $H_{0}$ because $\Omega_{m0}$ is a well-defined dimensionless cosmological parameter in cosmology. Using Eq.~\eqref{5.8} and \eqref{5.9} in \eqref{5.5}, we can obtain the Hubble function as
\begin{equation}
	H(z)=\frac{2H_{0}[\frac{\lambda_{0}}{H_{0}^{2}}+6\Omega_{m0}(1+z)^{3}]}{\frac{c_{1}}{H_{0}}+\sqrt{\left(\frac{c_{1}}{H_{0}}\right)^{2}+4\left(6\Omega_{m0}+\frac{\lambda_{0}}{H_{0}^{2}}-\frac{c_{1}}{H_{0}}\right)[\frac{\lambda_{0}}{H_{0}^{2}}+6\Omega_{m0}(1+z)^{3}] }}, \label{5.10}
\end{equation}
where $\Omega_{m0}$ denote the present value of the corresponding parameter and $H_{0}$ is the Hubble constant.\\
Now, Eqs.~\eqref{5.5} \& \eqref{5.6} can be rewritten as equivalent to Friedmann Equations.
\begin{equation}
	3H^{2}=\kappa^{2}\rho+\kappa^{2}\rho_{de}, \label{5.11}
\end{equation}
\begin{equation}
	2\dot{H}+3H^{2}=-\kappa^{2}p-\kappa^{2}p_{de}, \label{5.12}
\end{equation}
where $\rho_{de}$ and $p_{de}$ energy density and pressure come from geometrical modifications and are given, respectively, as
\begin{equation}
	\rho_{de}=\frac{1}{2\kappa^{2}}\left[\lambda_{0}-c_{1}H-6\lambda H^{2}\right], \label{5.13}
\end{equation}
\begin{equation}
	p_{de}=-\frac{1}{6\kappa^{2}}\left[3\lambda_{0}-3c_{1}H-18\lambda H^{2}-12\lambda \dot{H}-c_{1}\frac{\dot{H}}{H}\right], \label{5.14}
\end{equation}
Therefore, we derive the effective dark equation of state as
\begin{equation}
	\omega_{de}=-1-\frac{[c_{1}+12\lambda H](1+z)H'}{3\lambda_{0}-3c_{1}H-18\lambda H^{2}}. \label{5.15}
\end{equation}
Now, we can derive the deceleration parameter $q(z)$ from the Eq.~\eqref{5.10} as
\begin{equation}
	q(z)=-1+(1+z)\frac{H'}{H}  \label{5.16}
\end{equation}
where $H'=\frac{dH}{dz}$.

%====================================================
\subsection{Model-II}
%====================================================

In this model, we choose $u=c_{2}\frac{\ddot{a}}{a}$ as a function of second derivative of $a$ and $w=c_{3}\dot{a}$ as a function of first derivative of $a$ with $c_{2}$, $c_{3}$ as constants such that the energy conservation equation \eqref{5.3} is satisfied. However, there may be several such choices as per the concepts of $u, w$ (see \cite{ref71,ref79,ref80,ref81}). Now using these expressions of $u$ and $w$ in Eqs.~\eqref{5.1} \& \eqref{5.2}, we obtain the following simplified field equations
\begin{equation}
	-\frac{c_{2}}{2}\dot{H}+\frac{6(1+\lambda)-c_{2}}{2}H^{2}-\frac{\lambda_{0}}{2}=\kappa^{2}\rho, \label{5.17}
\end{equation}
\begin{equation}
	\frac{6(1+\lambda)-c_{2}}{3}\dot{H}+\frac{9(1+\lambda)-c_{2}}{3}H^{2}-\frac{\lambda_{0}}{2}=-\kappa^{2}p, \label{5.18}
\end{equation}
and the energy conservation equation \eqref{5.3} reduces to
\begin{equation}
	\dot{\rho}+3H(\rho+ p)=0, \label{5.19}
\end{equation}

Applying the third constraint on matter pressure as $p=0$, and using  Eq.~\eqref{5.8} in \eqref{5.17} and \eqref{5.18}, we get the Hubble function as
\begin{equation}
	H(z)=H_{0}\sqrt{1+\frac{2(6+6\lambda-c_{2})}{(1+\lambda)(12+12\lambda-c_{2})}\Omega_{m0}[(1+z)^{3}-1]},\,\label{5.20}
\end{equation}
where
\begin{equation}
	\frac{2\lambda_{0}(c_{2}+3+3\lambda)}{3(1+\lambda)(12+12\lambda-c_{2})}+\frac{2(6+6\lambda-c_{2})}{(1+\lambda)(12+12\lambda-c_{2})}=1.\,\label{5.21}
\end{equation}
Now, Eqs.~\eqref{5.17} \& \eqref{5.18} can be rewrite as equivalent to Friedmann Equations
\begin{equation}
	3H^{2}=\kappa^{2}\rho+\kappa^{2}\rho_{de}, \label{5.22}
\end{equation}
\begin{equation}
	2\dot{H}+3H^{2}=-\kappa^{2}p-\kappa^{2}p_{de}, \label{5.23}
\end{equation}
where effective energy density $\rho_{de}$ and pressure $p_{de}$ coming from geometrical modification in the Einstein's field equations, are given by respectively as
\begin{equation}
	\rho_{de}=\frac{1}{2\kappa^{2}}\left[\lambda_{0}+(c_{2}-6\lambda)H^{2}+c_{2}\dot{H}\right], \label{5.24}
\end{equation}
\begin{equation}
	p_{de}=-\frac{1}{6\kappa^{2}}\left[3\lambda_{0}+2(c_{2}-9\lambda)H^{2}+2(c_{2}-6\lambda)\dot{H}\right]. \label{5.25}
\end{equation}
Hence, the effective dark equation of state is derived as
\begin{equation}
	\omega_{de}=-1+\frac{c_{2}H^{2}-(c_{2}+12\lambda)(1+z)HH'}{3[\lambda_{0}+(c_{2}-6\lambda)H^{2}-c_{2}(1+z)HH']}. \label{5.26}
\end{equation}
Using Eq.~\eqref{5.20}, we derive the deceleration parameter as
\begin{equation}
	q(z)=-1+\frac{\frac{3(6+6\lambda-c_{2})}{(1+\lambda)(12+12\lambda-c_{2})}\Omega_{m0}(1+z)^{3}}{1+\frac{2(6+6\lambda-c_{2})}{(1+\lambda)(12+12\lambda-c_{2})}\Omega_{m0}[(1+z)^{3}-1]}. \label{5.27}
\end{equation}

%========================================================
\section{Observational Constraints}
%========================================================

    In this part, we utilize observational datasets to provide constraints on the model parameters in our derived model. To accomplish this, we utilize the emcee software, which is readily accessible at \cite{ref87}, to conduct a Monte Carlo Markov Chain (MCMC) analysis. This allows us to compare our generated model with observational datasets. The MCMC sampler restricts the cosmological and model parameters by varying their values within a plausible range of prior distributions and examining the resulting posterior distributions in the parameter space. In this section, we assess the compatibility between the solution in the model and the cosmic chronometer (CC) data and the Pantheon datasets. These datasets are related to the observed universe at a recent time frame.

%=========================================================
\subsection{Cosmic Chronometer (CC) Hubble datasets}
%=========================================================

The Hubble parameter holds significant importance for both theoretical and observational cosmologists as it is a crucial cosmological parameter for investigating the progression of the universe. Observed values for Hubble datasets $H(z)$ can be found for different redshifts $z$. To determine the optimal values for model parameters, taking into account the uncertainty range of redshift ($0.07\le z \le 1.965$), we employ a Monte Carlo Markov Chain (MCMC) simulation. This simulation allows us to compare the Hubble function derived from the field equations with the observed values of the 31 cosmic chronometer data points (referred to as Hubble data) \cite{ref88,ref89,ref90}. The values were determined using the differential ages (DA) of galaxies approach. To estimate the model parameters $H_{0}$, $\Omega_{m0}$, $c_{1}$, $c_{2}$ and $\lambda_{0}$, $\lambda$, we can minimize the $\chi^{2}$ function, which is equivalent to maximizing the likelihood function. The expression for the $\chi^{2}$ function is:
\begin{equation}\nonumber
	\chi_{CC}^{2}(\phi)=\sum_{i=1}^{i=N}\frac{[H_{ob}(z_{i})-H_{th}(\phi, z_{i})]^{2}}{\sigma_{H(z_{i})}^{2}}
\end{equation}
Where $N$ denotes the total amount of data, $H_{ob},~H_{th}$, respectively, the observed and hypothesized datasets of $H(z)$ and standard deviations are displayed by $\sigma_{H(z_{i})}$. Here $\phi=(H_{0}, \Omega_{m0}, c_{1}, \lambda_{0})$ for Model-I and for Model-II $\phi=(H_{0}, \Omega_{m0}, \lambda, c_{2})$.\\

%%%%%%%%%%%%%%%%%%%%%%%%%%%%%%%%%%%%%%%%%%%%%%%%%%%%%%%%%%%%
%%%%%%%%%%%%%%%%%%%%%%%%%%%%%%%%%%%% Figure 1
%%%%%%%%%%%%%%%%%%%%%%%%%%%%%%%%%%%%%%%%%%%%%%%%%%%%%%%%%%%%
\begin{figure}[H]
	\centering
	\includegraphics[width=10cm,height=10cm,angle=0]{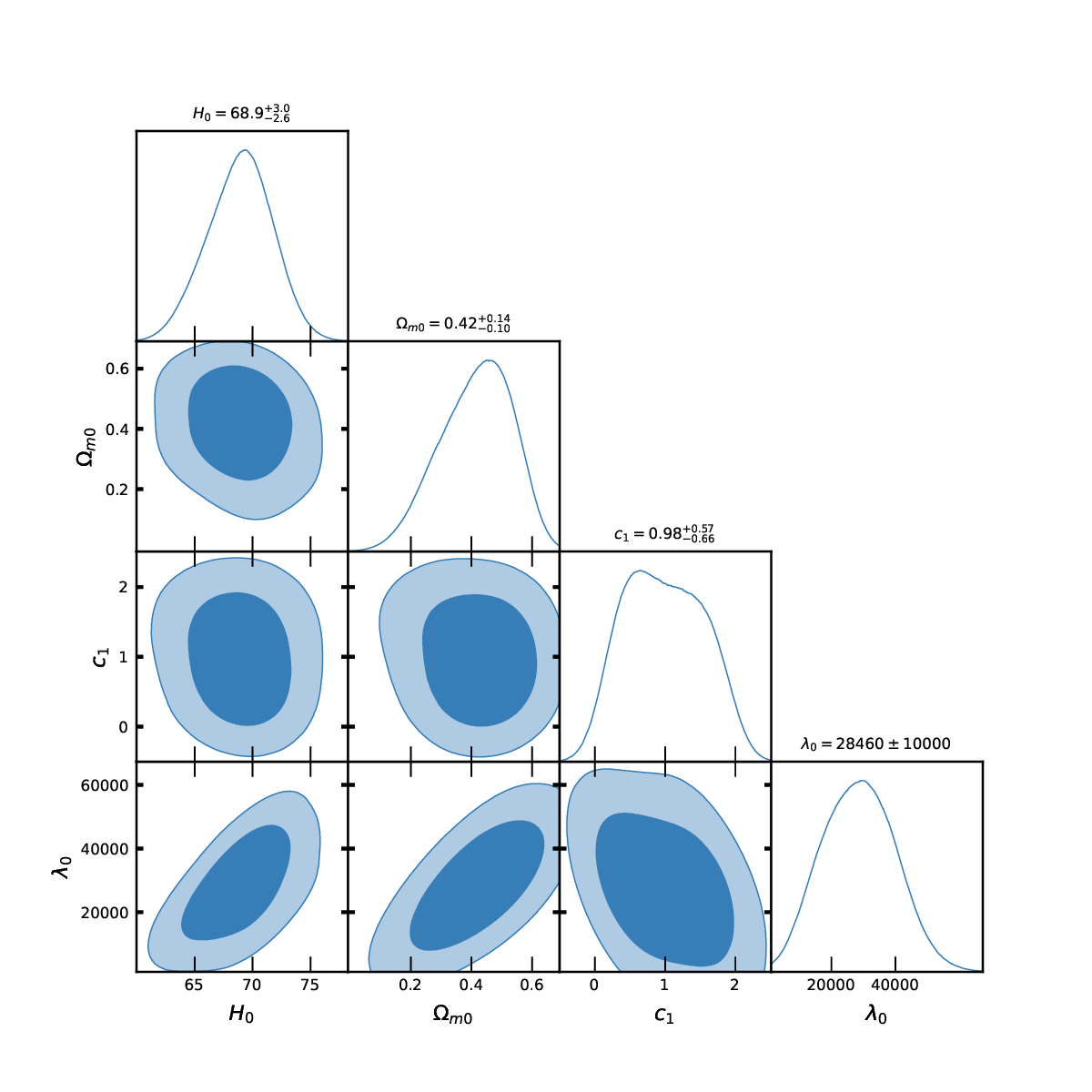}
	\caption{The contour plots of $H_{0}, \Omega_{m0}, c_{1}, \lambda_{0}$ at $1-\sigma$ and $2-\sigma$ confidence level in MCMC analysis of CC datasets for Model-I.}
\end{figure}
%%%%%%%%%%%%%%%%%%%%%%%%%%%%%%%%%%%%%%%%%%%%%%%%%%%%%%%%%%%%%%%%%%

%%%%%%%%%%%%%%%%%%%%%%%%%%%%%%%%%%%%%%%%%%%%%%%%%%%%%%%%%%%%
%%%%%%%%%%%%%%%%%%%%%%%%%%%%%%%%%%%% Figure 2
%%%%%%%%%%%%%%%%%%%%%%%%%%%%%%%%%%%%%%%%%%%%%%%%%%%%%%%%%%%%
\begin{figure}[H]
	\centering
	\includegraphics[width=10cm,height=10cm,angle=0]{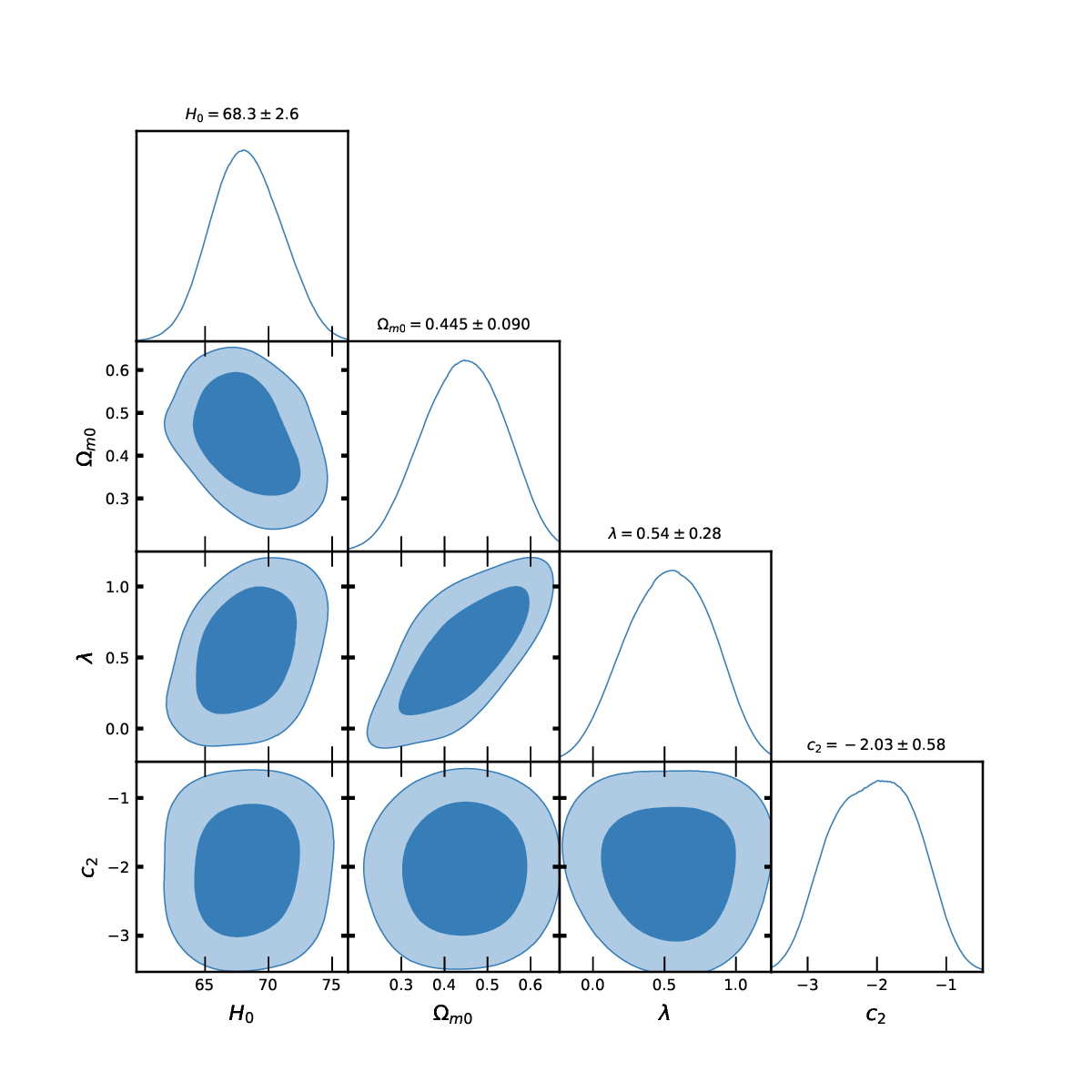}
	\caption{The contour plots of $H_{0}, \Omega_{m0}, \lambda, c_{2}$ at $1-\sigma$ and $2-\sigma$ confidence level in MCMC analysis of CC datasets for Model-II.}
\end{figure}
%%%%%%%%%%%%%%%%%%%%%%%%%%%%%%%%%%%%%%%%%%%%%%%%%%%%%%%%%%%%%%%%%%

	%%%%%%%%%%%%%%%%%%%%%%%%%%%%%%%%%%%%%%%%%%%%%%%%%%%%%%%%%%%%
%%%%%%%%%%%%%%%%%%%%%%%%%%%%%%%%%%%% Figure 3
%%%%%%%%%%%%%%%%%%%%%%%%%%%%%%%%%%%%%%%%%%%%%%%%%%%%%%%%%%%%
\begin{figure}[H]
	\centering
	\includegraphics[width=8cm,height=7cm,angle=0]{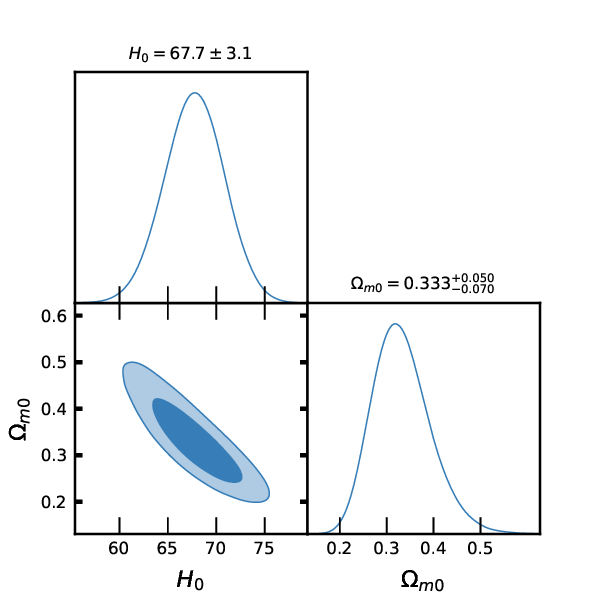}
	\caption{The contour plots of $H_{0}, \Omega_{m0}$ at $1-\sigma$ and $2-\sigma$ confidence level in MCMC analysis of CC datasets for $\Lambda$CDM.}
\end{figure}
%%%%%%%%%%%%%%%%%%%%%%%%%%%%%%%%%%%%%%%%%%%%%%%%%%%%%%%%%%%%%%%%%%
%%%%%%%%%%%%%%%%%%%%%%%%%%%%%%%%%%%%%%%%%%%%%%%%%%%%%%%%%%%%%%%%%%%%
\begin{table}[H]
	\centering
	\begin{tabular}{|c|c|c|c|}
		\hline
		% after \\: \hline or \cline{col1-col2} \cline{col3-col4} ...
		Model    &	Parameter       & Prior            & Value   \\
		\hline
		         &$H_{0}$        & $(40, 100)$      & $68.9_{-2.6}^{+3.0}$\\
                 &$\Omega_{m0}$  & $(0, 0.6)$       & $0.42_{-0.10}^{+0.14}$\\
         Model-I &$\lambda_{0}$  & $(10000, 60000)$ & $28460\pm10000$\\
                 &$c_{1}$        & $(0, 2)$           & $0.98_{-0.66}^{+0.57}$\\
                 &$\chi_{min}^{2}$     & --               & $14.493$\\
		\hline
		         &$H_{0}$        & $(40, 100)$      & $68.3\pm2.6$\\
                 &$\Omega_{m0}$  & $(0, 0.6)$       & $0.445\pm0.090$\\
        Model-II &$\lambda$      & $(0, 1)$          & $0.54\pm0.28$\\
                 &$c_{2}$        & $(-3, 0)$         & $-2.03\pm0.58$\\
                 &$\chi_{min}^{2}$     & --               & $14.494$\\
		\hline
				 &$H_{0}$        & $(40, 100)$      & $67.7\pm3.1$\\
   $\Lambda$CDM  &$\Omega_{m0}$  & $(0, 1)$         & $0.333_{-0.07}^{+0.05}$\\
                 &$\chi_{min}^{2}$     & --               & $14.494$\\
		\hline
	\end{tabular}
	\caption{The MCMC Results in $H(z)$ datasets analysis.}\label{T1}
\end{table}
%%%%%%%%%%%%%%%%%%%%%%%%%%%%%%%%%%%%%%%%%%%%%%%%%%%%%%%%%%%%%%%%%%%%%%%%%%%%

We employ Bayesian statistical analysis for Markov chain Monte Carlo (MCMC) simulation to calibrate the CC datasets. To achieve this, we use the emcee package developed by Foreman-Mackey \textit{et al.} \cite{ref87}. We have reduced the chi-squared statistic, $\chi_{CC}^{2}(\phi)$, in order to find the best values for the model's parameters. Table 1 presents the values. \\

%============================================================
\subsection{Distance Modulus $\mu(z)$}
%============================================================

The correlation between luminosity distance and redshift is a fundamental observational method employed to monitor the progression of the cosmos. When calculating the luminosity distance ($D_{L}$) in relation to the cosmic redshift ($z$), the expansion of the universe and the redshift of light from distant bright objects are factored in. It is given as
\begin{equation}\label{6.1}
	D_{L}=a_{0} r (1+z),
\end{equation}
where the radial coordinate of the source $r$, is established by
\begin{equation}\label{6.2}
	r  =  \int^r_{0}dr = \int^t_{0}\frac{cdt}{a(t)} = \frac{1}{a_{0}}\int^z_0\frac{cdz'}{H(z')},
\end{equation}
where we have used $ dt=dz/\dot{z}, \dot{z}=-H(1+z)$.\\
Consequently, the following formula determines the luminosity distance:
\begin{equation}\label{6.3}
	D_{L}=c(1+z)\int^z_0\frac{dz'}{H(z')}.
\end{equation}
Supernovae (SNe) are commonly employed by researchers as standard candles to investigate the pace of cosmic expansion using the reported apparent magnitude ($m_{o}$). The surveys on supernovae that discovered several types of supernovae of varying magnitudes resulted in the creation of the Pantheon sample SNe datasets, comprising $1048$ data points within the range of $0.01$ to $2.26$ for the variable $z$. The theoretical apparent magnitude ($m$) of these standard candles is precisely defined as \cite{ref91}.
\begin{equation}\label{6.4}
	m(z)=M+ 5~\log_{10}\left(\frac{D_{L}}{Mpc}\right)+25.
\end{equation}
where $M$ represents the absolute magnitude. The luminosity distance is quantified in units of distance. The Hubble-free luminosity distance ($d_{L}$) can be expressed as $d_{L}\equiv\frac{H_{0}}{c}D_{L}$, where $D_{L}$ is a dimensionless quantity based on $D_{L}$. Therefore, we can express $m(z)$ in a simplified form as shown below
\begin{equation}\label{6.5}
	m(z)=M+5\log_{10}{d_{L}}+5\log_{10}\left(\frac{c/H_{0}}{Mpc}\right)+25.
\end{equation}
The equation provided allows for the observation of the degeneracy between $M$ and $H_{0}$, which remains constant in the $\Lambda$CDM background \cite{ref91,ref92}. By redefining, we can combine these deteriorated parameters.
\begin{equation}\label{6.6}
	\mathcal{M}\equiv M+5\log_{10}\left(\frac{c/H_{0}}{Mpc}\right)+25.
\end{equation}
The dimensionless parameter $\mathcal{M}$ is defined by the equation $\mathcal{M}=M-5\log_{10}(h)+42.39$, where $H_{0}$ is equal to $h\times100 Km/s/Mpc$. In the Markov Chain Monte Carlo (MCMC) analysis, we utilize this parameter in conjunction with the appropriate $\chi^{2}$ value for the Pantheon data, as provided in \cite{ref93}.
\begin{equation}\label{6.7}
	\chi^{2}_{P}=V_{P}^{i}C_{ij}^{-1}V_{P}^{j}
\end{equation}
The expression $V_{P}^{i}$ is defined as the difference between $m_{o}(z_{i})$ and $m(z)$. The matrix $C_{ij}$ is the inverse of the covariance matrix, and the value of $m(z)$ is determined by Equation \eqref{6.5}.

%%%%%%%%%%%%%%%%%%%%%%%%%%%%%%%%%%%%%%%%%%%%%%%%%%%%%%%%%%%%
%%%%%%%%%%%%%%%%%%%%%%%%%%%%%%%%%%%% Figure 4
%%%%%%%%%%%%%%%%%%%%%%%%%%%%%%%%%%%%%%%%%%%%%%%%%%%%%%%%%%%%
\begin{figure}[H]
	\centering
	\includegraphics[width=10cm,height=10cm,angle=0]{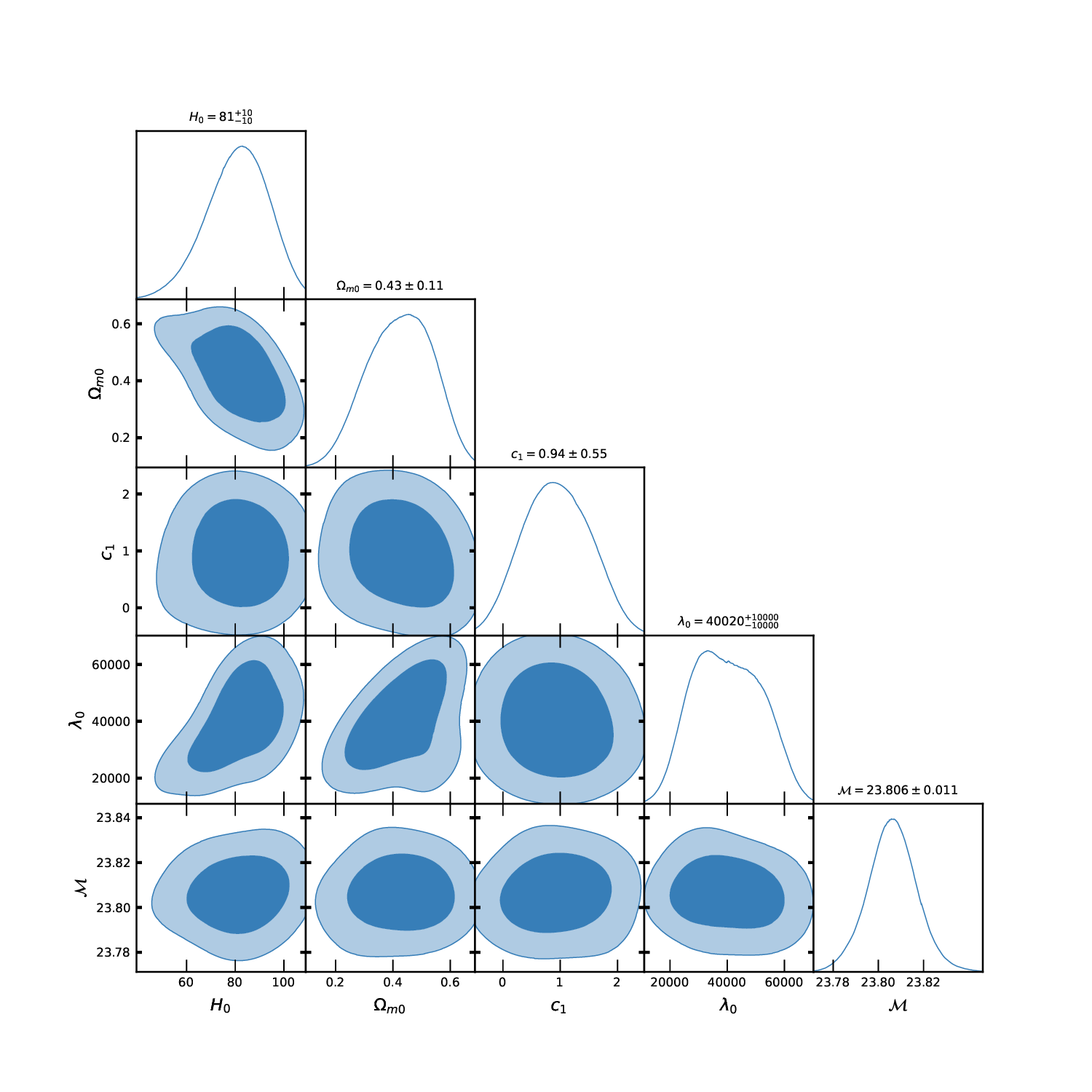}
	\caption{The contour plots of $H_{0}, \Omega_{m0}, c_{1}, \lambda_{0}, \mathcal{M}$ at $1-\sigma$ and $2-\sigma$ confidence level in MCMC analysis of Pantheon SNe Ia datasets for Model-I.}
\end{figure}
%%%%%%%%%%%%%%%%%%%%%%%%%%%%%%%%%%%%%%%%%%%%%%%%%%%%%%%%%%%%%%%%%%

%%%%%%%%%%%%%%%%%%%%%%%%%%%%%%%%%%%%%%%%%%%%%%%%%%%%%%%%%%%%
%%%%%%%%%%%%%%%%%%%%%%%%%%%%%%%%%%%% Figure 5
%%%%%%%%%%%%%%%%%%%%%%%%%%%%%%%%%%%%%%%%%%%%%%%%%%%%%%%%%%%%
\begin{figure}[H]
	\centering
	\includegraphics[width=10cm,height=10cm,angle=0]{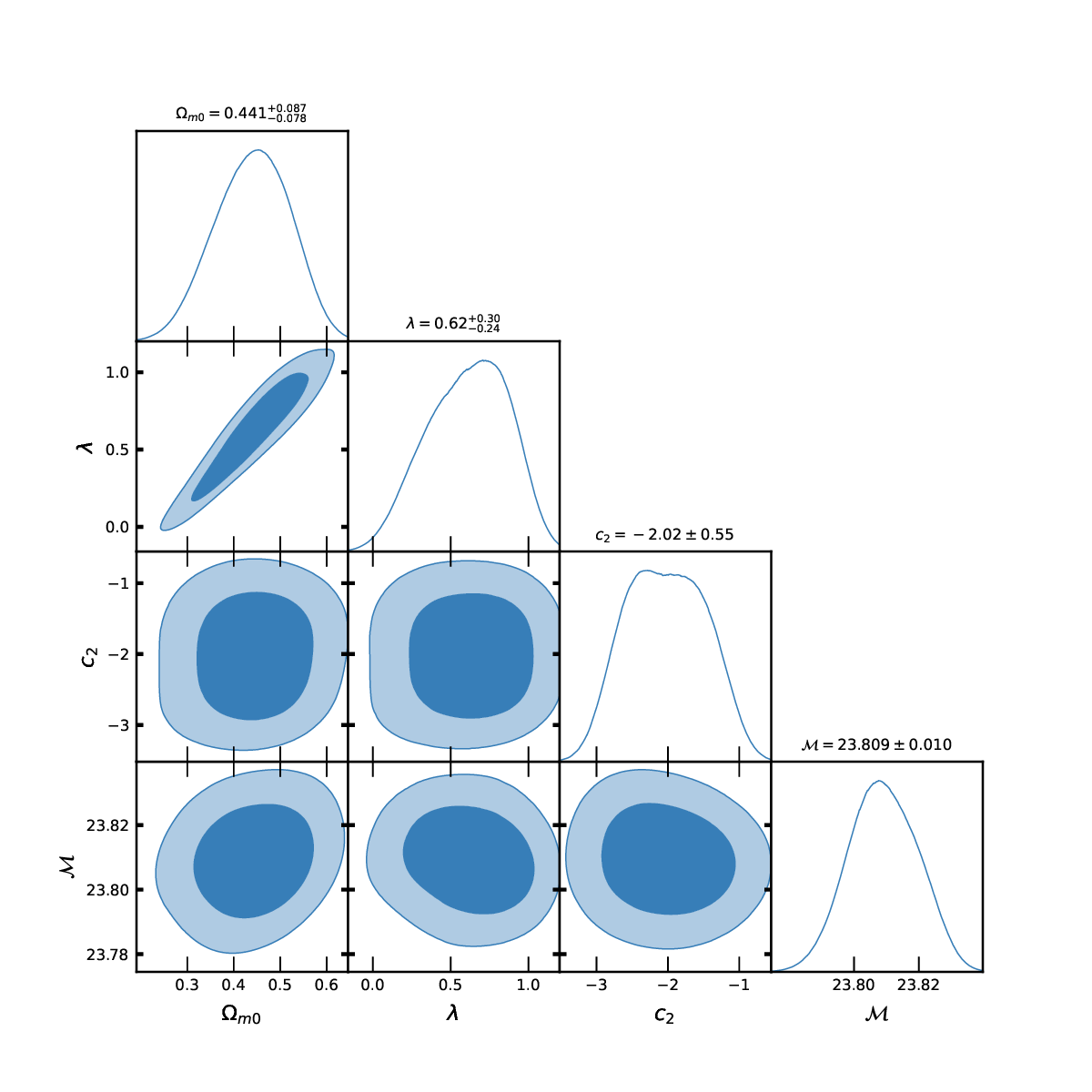}
	\caption{The contour plots of $H_{0}, \Omega_{m0}, \lambda, c_{2}, \mathcal{M}$ at $1-\sigma$ and $2-\sigma$ confidence level in MCMC analysis of Pantheon SNe Ia datasets for Model-II.}
\end{figure}
%%%%%%%%%%%%%%%%%%%%%%%%%%%%%%%%%%%%%%%%%%%%%%%%%%%%%%%%%%%%%%%%%%

%%%%%%%%%%%%%%%%%%%%%%%%%%%%%%%%%%%%%%%%%%%%%%%%%%%%%%%%%%%%
%%%%%%%%%%%%%%%%%%%%%%%%%%%%%%%%%%%% Figure 6
%%%%%%%%%%%%%%%%%%%%%%%%%%%%%%%%%%%%%%%%%%%%%%%%%%%%%%%%%%%%
\begin{figure}[H]
	\centering
	\includegraphics[width=8cm,height=7cm,angle=0]{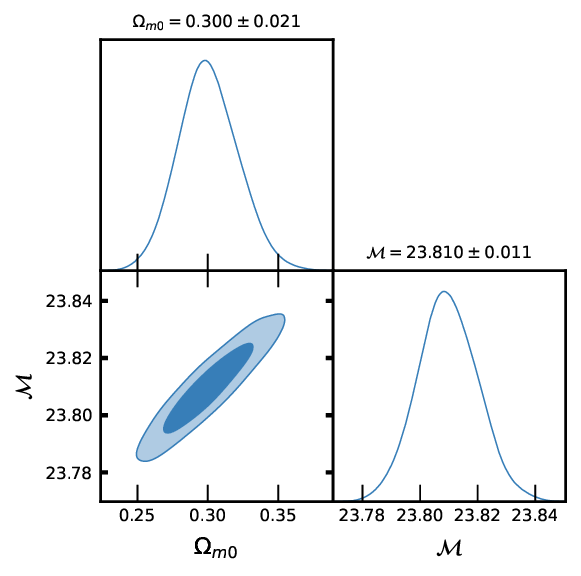}
	\caption{The contour plots of $H_{0}, \Omega_{m0}, \mathcal{M}$ at $1-\sigma$ and $2-\sigma$ confidence level in MCMC analysis of Pantheon SNe Ia datasets for $\Lambda$CDM.}
\end{figure}
%%%%%%%%%%%%%%%%%%%%%%%%%%%%%%%%%%%%%%%%%%%%%%%%%%%%%%%%%%%%%%%%%%
%%%%%%%%%%%%%%%%%%%%%%%%%%%%%%%%%%%%%%%%%%%%%%%%%%%%%%%%%%%%%%%%%%%%
\begin{table}[H]
	\centering
	\begin{tabular}{|c|c|c|c|}
		\hline
		% after \\: \hline or \cline{col1-col2} \cline{col3-col4} ...
		Model    &	Parameter    & Prior            & Value   \\
		\hline
		         &$H_{0}$        & $(40, 100)$      & $81.0\pm10$\\
		         &$\Omega_{m0}$  & $(0, 0.6)$       & $0.43\pm0.11$\\
		 Model-I &$\lambda_{0}$  & $(10000, 60000)$ & $40020\pm10000$\\
		         &$\mathcal{M}$  & $(23, 24)$       & $23.806\pm0.011$\\
		         &$c_{1}$        & $(0, 2)$          & $0.94\pm0.55$\\
		         &$\chi_{min}^{2}$& --               & $1026.670$\\
		\hline
		         &$\Omega_{m0}$  & $(0, 0.6)$       & $0.441_{-0.078}^{+0.087}$\\
                 &$\lambda$      & $(0, 1)$         & $0.62_{-0.24}^{+0.30}$\\
        Model-II &$\mathcal{M}$  & $(23, 24)$       & $23.809\pm0.010$\\
                 &$c_{2}$        & $(-3, 0)$        & $-2.02\pm0.55$\\
                 &$\chi_{min}^{2}$     & --               & $1026.671$\\
        \hline
                 &$\Omega_{m0}$  & $(0, 1)$         & $0.300\pm0.021$\\
    $\Lambda$CDM &$\mathcal{M}$  & $(23, 24)$       & $23.810\pm0.011$\\
                 &$\chi_{min}^{2}$     & --               & $1026.671$\\
		\hline
	\end{tabular}
	\caption{The MCMC Results in Pantheon SNe Ia datasets analysis.}\label{T2}
\end{table}
%%%%%%%%%%%%%%%%%%%%%%%%%%%%%%%%%%%%%%%%%%%%%%%%%%%%%%%%%%%%%%%%%%%%%%%%%%%%

%===========================================================
\subsection*{Statistical Analysis:}
%===========================================================

This section examines several cosmological theories using the Akaike Information Criterion (AIC) and the Bayesian Information Criterion (BIC). Furthermore, we calculate the reduced chi-squared value by employing the method ($\chi_{red}^{2}=\chi^{2}_{min}/dof$), where ``dof" denotes the degrees of freedom. We commonly determine the degrees of freedom by subtracting the number of fitted parameters from the number of data points used. For elucidation purposes, however, it is advisable to exclusively employ the $\chi^{2}_{min}/dof$ metric, as the degrees of freedom may not be apparent for models that do not exhibit linearity in relation to the independent parameters \cite{ref94}. The AIC criteria, which are based on information theory, act as an estimator of asymptotically unbiased Kullback-Leibler information. The AIC criteria can be approximated using the formula stated in references \cite{ref95,ref96}, assuming Gaussian errors.\\

\begin{equation}\label{6.8}
	AIC=-2\ln(\mathcal{L}_{max})+2n+\frac{2n(n+1)}{N-n-1}
\end{equation}
The symbol $\mathcal{L}_{max}$ denotes the maximum likelihood of the dataset(s) being analyzed. The variable $N$ reflects the total number of data points used in the analysis, whereas $n$ represents the number of fitted parameters. Maximizing the likelihood function is synonymous with minimizing the $\chi^{2}$ value. When $N$ is a big value, it is clear that this expression produces the original version of AIC, which can be approximated as $AIC\backsimeq-2\ln(\mathcal{L}_{max})+2n$. As stated in the discussion in \cite{ref97}, the utilization of the modified AIC criterion is usually considered the most effective strategy. The BIC criteria is a Bayesian evidence estimator, and it is cited by \cite{ref95,ref96,ref97}.

\begin{equation}\label{6.9}
	BIC=-2\ln(\mathcal{L}_{max})+n\ln(N)
\end{equation}

Our goal is to organize the models according to their ability to accurately correspond to the given data, taking into account a set of scenarios that portray the same kind of occurrence. To determine the disparity in the information criteria (IC) value for a given collection of models, we employ the two ICs mentioned before. The expression $\Delta IC_{model}=IC_{model}-IC_{min}$ represents the difference between a model's IC value ($IC_{model}$) and the model's IC value with the lowest IC value ($IC_{min}$). In order to assess the appropriateness of each model, we employ the Jeffreys scale \cite{ref98}. Specifically, when the value of $\Delta IC$ is less than or equal to 2, it signifies that the data provides significant evidence in favor of the most preferred model. When the difference between IC values is between 2 and 6, it indicates a considerable amount of disagreement between the two models. Finally, when the difference in IC (Information Criterion) is greater than or equal to 10, it indicates a significant degree of tension between the models \cite{ref71}.\\

Our approach incorporates two distinct datasets: the cosmic chronometer (Hubble data) points and the Pantheon SNe Ia datasets. The model parameters for our derived models have been fitted by minimizing the $\chi^{2}$ value. The resulting values of $\chi_{min}^{2}$ are displayed in Tables 1 and 2, respectively. For models I and II, we calculated the minimum chi-square value ($\chi_{min}^{2}=14.493, 14.494$), respectively, using CC datasets while for $\Lambda$CDM $\chi^{2}=14.494$. We also determined the values of AIC and BIC, which are presented in Table 3, along with the difference from the best-fitted model ($\Delta IC_{model}=IC_{model}-IC_{min}$). The total number of data points is $N=31$ and the number of parameters is $n=4$ for Models I and II while for $\Lambda$CDM $n=2$.
%%%%%%%%%%%%%%%%%%%%%%%%%%%%%%%%%%%%%%%%%%%%%%%%%%%%%%%%%%%%%%%%%%%%%	
\begin{table}[H]
	\centering
	\begin{tabular}{|c|c|c|c|c|}
		\hline
		% after \\: \hline or \cline{col1-col2} \cline{col3-col4} ...
		
		Model        & AIC        & $\Delta$AIC  & BIC       & $\Delta$BIC   \\
		\hline
		Model-I      & $24.032$  & $5.110$     & $28.229$ & $6.868$\\
		Model-II     & $24.032$  & $5.110$     & $28.230$ & $6.868$\\
		$\Lambda$CDM & $18.922$  & $0$          & $21.362$ & $0$\\				
		\hline
	\end{tabular}
	\caption{The information criteria AIC and BIC for the examined cosmological models along cosmic chronometer datasets.}\label{T3}
\end{table}
%%%%%%%%%%%%%%%%%%%%%%%%%%%%%%%%%%%%%%%%%%%%%%%%%%%%%%%%%%%%%%%%%%%%%%%%%%%%
We used $\chi^{2}=1026.670$, $N=1048$ and $n=5$ for Model-I, $\chi^{2}=1026.671$, $N=1048$ and $n=4$ for Model-II while for $\Lambda$CDM, we used $\chi^{2}=1026.671$, $N=1048$ and $n=2$ to get the AIC and BIC values for the Pantheon SNe Ia datasets. The AIC and BIC values are shown below in Table 4, along with the difference from the best-fitting model, which is $\Delta IC_{model}=IC_{model}-IC_{min}$.
%%%%%%%%%%%%%%%%%%%%%%%%%%%%%%%%%%%%%%%%%%%%%%%%%%%%%%%%%%%%%%%%%%%%%	
\begin{table}[H]
	\centering
	\begin{tabular}{|c|c|c|c|c|}
		\hline
		% after \\: \hline or \cline{col1-col2} \cline{col3-col4} ...
		
		Model        & AIC          & $\Delta$AIC  & BIC         & $\Delta$BIC   \\
		\hline
		Model-I      & $1036.728$  & $6.046$    & $1061.443$ & $20.863$\\
		Model-II     & $1034.709$  & $4.027$     & $1054.489$ & $13.909$\\
		$\Lambda$CDM & $1030.682$  & $0$          & $1040.580$ & $0$\\				
		\hline
	\end{tabular}
	\caption{The information criteria AIC and BIC for the examined cosmological models, along Pantheon SNe Ia datasets.}\label{T4}
\end{table}
%%%%%%%%%%%%%%%%%%%%%%%%%%%%%%%%%%%%%%%%%%%%%%%%%%%%%%%%%%%%%%%%%%%%%%%%%%%%

%====================================================================
\section{Result discussions}
%====================================================================

Based on the findings presented in the previous section, we proceed to examine FLRW cosmological models under metric-affine $F(R,Q)$ gravity from an observational perspective. It is important to emphasize that the models stated above have some parameters that are free to be determined. These parameters include $H_{0}$, $\Omega_{m0}$, $\lambda$, $\lambda_{0}$, $c_{1}$ and $c_{2}$. However, in the case of concordance cosmology, the only free parameters are $H_{0}$, $\Omega_{m0}$. To enhance convenience, we compile the acquired outcomes in Tables 1 and 2. In addition, we provide contour plots for Model I and Model II in Figures 1, 2, 4, and 5, respectively. In addition, we examined the concordance model, namely the $\Lambda$CDM model, in order to compare and establish a standard for evaluation. Figure 3 and 6 depict the contour plots for $\Lambda$CDM corresponding to two observational datasets CC and Pantheon SNe Ia, respectively.\\

In MCMC analysis of CC $H(z)$ datasets and Pantheon SNe Ia datasets for Model-I and Model-II, Figures 1, 2, 4, and 5 show the contour plots of $H_{0}, \Omega_{m0}, \lambda, \lambda_{0}, c_{1}$ and $c_{2}$ at $1-\sigma$ and $2-\sigma$ confidence levels. Table 1 and Table 2 display the estimated values of cosmological parameters. The dimensionless parameter $\lambda$ is constrained to an interval around $0$, which includes the $\Lambda$CDM paradigm, which was expected since, as we discussed above, a realistic modified gravity should be a small deviation from general relativity. Nevertheless, note that in both Model-I and Model-II, the $\lambda$-contours are slightly shifted towards positive values. To relax the degeneracy between the parameters $\lambda$, $\lambda_{0}$ and $c_{1}, c_{2}$, we have eliminated $\lambda$ in Model-I and $\lambda_{0}$ in Model-II. Recently, \cite{ref71} estimated the value of $\lambda=0.491_{-0.533}^{+0.387}, 0.537_{-0.550}^{+0.403}$, respectively, in two different models. The current values of model parameter $\lambda_{0}$ of dimension $H_{0}^{2}$ are estimated as $\lambda_{0}=28460\pm10000, 40020\pm10000$ along two datasets. We have estimated the approximate values of $\lambda$ as $\lambda=0.417_{-0.398}^{+0.380}, 0.445_{-0.133}^{+0.099}$ for Model-I and $\lambda=0.54\pm0.28, 0.62_{-0.24}^{+0.30}$ for Model-II, along two observational datasets, respectively.\\

In the context of estimated values of $\Omega_{m0}$, we observe that Model-I and II give a rather large value due to the degeneracy with $\lambda$, $\lambda_{0}$ and $c_{1}$, $c_{2}$, while in $\Lambda$CDM, this is not the case. Concerning the Hubble constant $H_{0}$, for model-I, we find that $68.9_{-2.6}^{+3.0}, 81.0\pm10$ Km/s/Mpc, while for model-II, we get $68.3\pm2.6$ Km/s/Mpc, along two datasets, respectively. For the $\Lambda$CDM, we obtain the value of the Hubble constant as $H_{0}=67.7\pm3.1$ Km/s/Mpc. The values of $H_{0}$ obtained in our estimation for Model-I and Model-II are large in comparison to $\Lambda$CDM, due to the degeneracy with other parameters of the models. Recently, the present value of the Hubble constant was measured as $H_{0}=69.8\pm1.3~Km s^{-1} Mpc^{-1}$ in \cite{ref99}, and $H_{0}=69.7\pm1.2~Km s^{-1} Mpc^{-1}$ was estimated in \cite{ref100}. This number is found to be $H_{0}=66.6\pm1.6~Km s^{-1} Mpc^{-1}$ by looking at a lot of observational data in \cite{ref101}. It is also found to be $H_{0}=65.8\pm3.4~Km s^{-1} Mpc^{-1}$ by looking at \cite{ref102,ref103}. The Hubble constant has been measured as $H_{0}=69.6\pm0.8~Km s^{-1} Mpc^{-1}$ in \cite{ref104}, $H_{0}=67.4_{-3.2}^{+4.1}~Km s^{-1} Mpc^{-1}$ in \cite{ref105}, $H_{0}=69_{-2.8}^{+2.9}~Km s^{-1} Mpc^{-1}$ in \cite{ref106}, and most recently, $H_{0}=68.81_{-4.33}^{+4.99}~Km s^{-1}Mpc^{-1}$ in \cite{ref107}. In 2018 \cite{ref108}, the Hubble constant was estimated by the Plank Collaboration to be $H_{0}=67.4\pm0.5$ km/s/Mpc, whereas in 2021, $H_{0}=73.2\pm1.3$ km/s/Mpc was determined in \cite{ref109}. Recently, \cite{ref110} has estimated the value Hubble constant as $H_{0}=69.504_{-0.141}^{+0.149}$ Km/s/Mpc, and \cite{ref111} estimates the value $H_{0}=68_{-2.0}^{+2.3}$ Km/s/Mpc, while in \cite{ref112}, the value of Hubble constant is reported as $H_{0}=73.5\pm1.1$ Km/s/Mpc for Pantheon+ datasets. When compared to previous results, the outcomes of our models I and II for $H_{0}$ are consistent with observational datasets. For two different models, we estimated the value of parameter $\mathcal{M}=23.806\pm0.011, 23.809\pm0.010$, while for $\Lambda$CDM, it is found as $\mathcal{M}=23.810\pm0.011$. Recently, \cite{ref113} estimated the value of $\mathcal{M}=23.809\pm0.013$.\\

Equations \eqref{5.15} and \eqref{5.26} represent the expressions of $\omega_{de}$ for Models I and II, respectively. The variations of $\omega_{de}$ over redshift $z$ are depicted in figures 7a and 7b, respectively, for Models I and II. From figure 7a, one can see that $\omega_{de}>-\frac{1}{3}$ for $z_{t}>0.681, 0.678$ for two datasets, respectively, which corresponds to a decelerating expansion phase of the universe, while $\omega_{de}<-\frac{1}{3}$ over $-1 \le z < 0.681, 0.678$ corresponds to the accelerating expansion phase of the universe. The lines $z_{t}=0.681, 0.678$ show the phase transition line of the expanding universe with $\omega_{de}=-\frac{1}{3}$. The present estimated values of $\omega_{de}=-1.214_{-0.351}^{+0.206}$ for CC datasets and $\omega_{de}=-1.233_{-0.182}^{+0.121}$ for Pantheon datasets are $\omega_{de}\to-1$ as $z\to-1$. Figure 7b shows that the dark energy EoS parameter $\omega_{de}>-\frac{1}{3}$ for transition redshifts $z_{t}=0.626, 0.677$ along two datasets used for Model-II, and $\omega_{de}<-\frac{1}{3}$ over $-1\le z <0.626, 0.677$ along the same two datasets. The present value of $\omega_{de}$ for Model-II is estimated as $\omega_{de}=-0.679_{-0.006}^{+0.007}$ for CC datasets, and $\omega_{de}=-0.694_{-0.009}^{+0.006}$ along Pantheon datasets, which corresponds to the accelerating phase of the expanding universe. Also, we have measured the late-time values of the EoS parameter as $\omega_{de}=-0.872_{-0.053}^{+0.080}, -0.883_{-0.047}^{+0.099}$ along two datasets, respectively. Thus, both Models I and II are transit phase (decelerating to accelerating) expanding universe models.

%%%%%%%%%%%%%%%%%%%%%%%%%%%%%%%%%%%%%%%%%%%%%%%%%%%%%%%%%%%%
%%%%%%%%%%%%%%%%%%%%%%%%%%%%%%%%%%%% Figure 7
%%%%%%%%%%%%%%%%%%%%%%%%%%%%%%%%%%%%%%%%%%%%%%%%%%%%%%%%%%%%
\begin{figure}[H]
	\centering
	a.\includegraphics[width=8cm,height=7cm,angle=0]{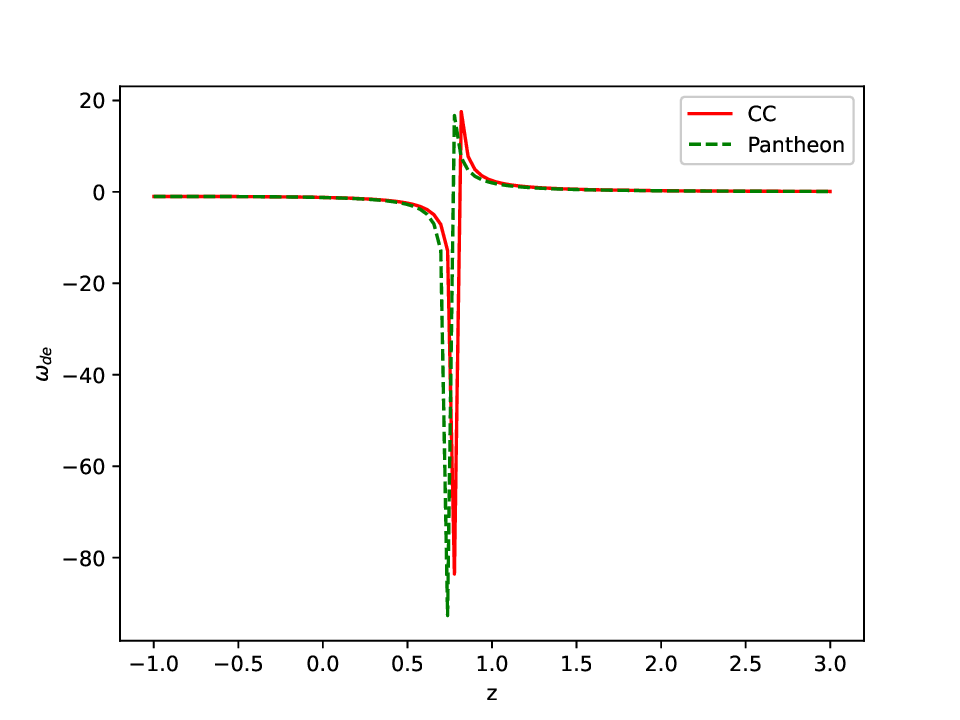}
	b.\includegraphics[width=8cm,height=7cm,angle=0]{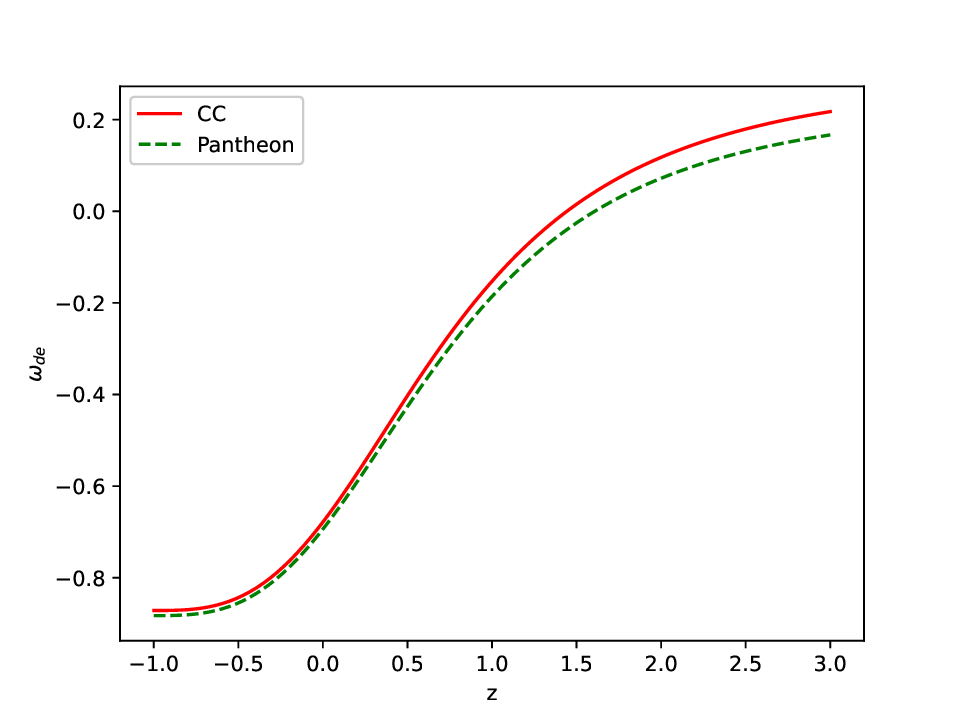}
	\caption{The variation of dark energy EoS parameter $\omega_{de}$ over redshift $z$ for Model-I and Model-II, respectively.}
\end{figure}
%%%%%%%%%%%%%%%%%%%%%%%%%%%%%%%%%%%%%%%%%%%%%%%%%%%%%%%%%%%%%%%%%%
%%%%%%%%%%%%%%%%%%%%%%%%%%%%%%%%%%%%%%%%%%%%%%%%%%%%%%%%%%%%
%%%%%%%%%%%%%%%%%%%%%%%%%%%%%%%%%%%% Figure 8
%%%%%%%%%%%%%%%%%%%%%%%%%%%%%%%%%%%%%%%%%%%%%%%%%%%%%%%%%%%%
\begin{figure}[H]
	\centering
	a.\includegraphics[width=8cm,height=7cm,angle=0]{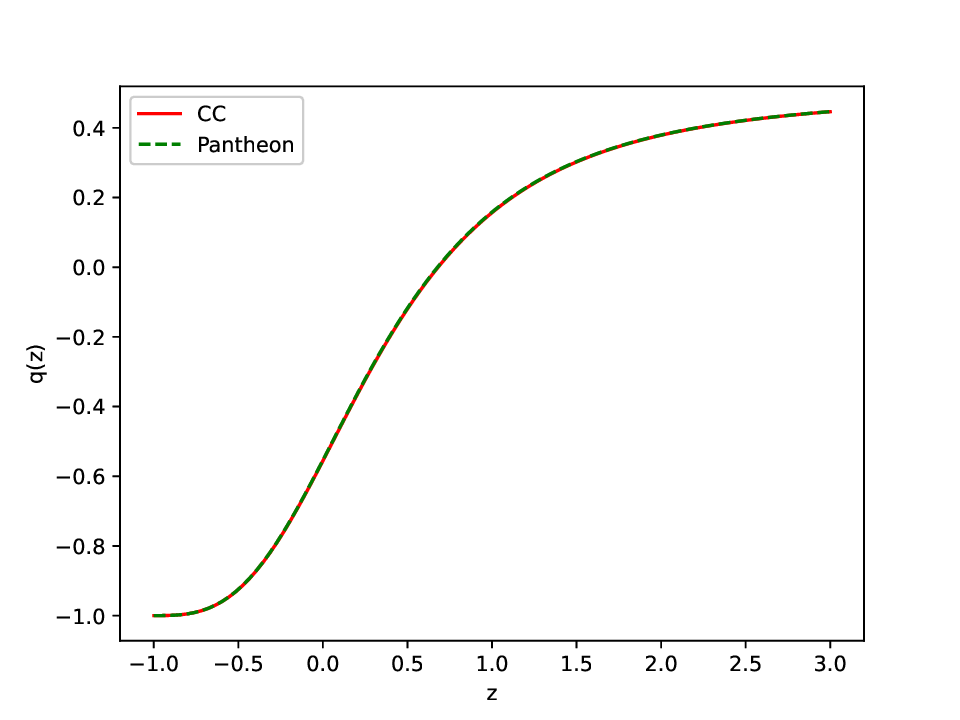}
    b.\includegraphics[width=8cm,height=7cm,angle=0]{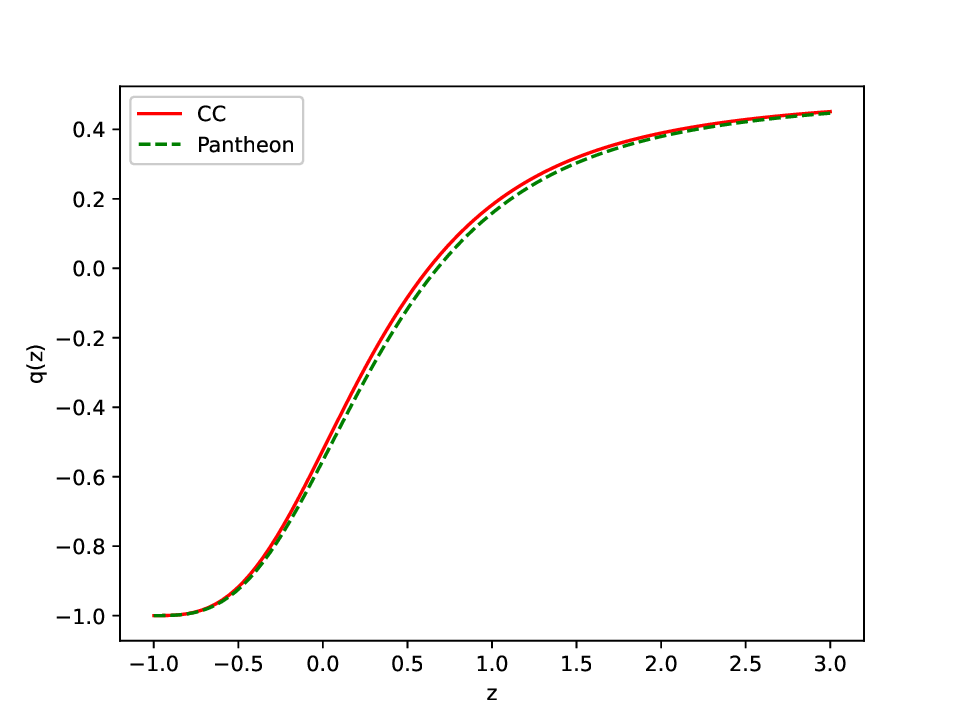}
	\caption{The variation of deceleration parameter $q(z)$ over redshift $z$ for Model-I and Model-II, respectively.}
\end{figure}
%%%%%%%%%%%%%%%%%%%%%%%%%%%%%%%%%%%%%%%%%%%%%%%%%%%%%%%%%%%%%%%%%%
The expressions for the deceleration parameter $q(z)$ are represented by the equations \eqref{5.16} and \eqref{5.27}, respectively, for Models I and II. Figures 8a and 8b, respectively, depict the geometrical evolution of $q(z)$ for Models I and II. From figures 8a and 8b, it is clear that our two derived models are transit-phase universe models, which are decelerating in the past and accelerating in late-time scenarios. The transition redshift is measured as $z_{t}=0.681, 0.678$ for the Model-I and $z_{t}=0.626, 0.677$ for the Model-II, along with two datasets, CC and Pantheon, respectively, which are consistent with recent observed values. The present value of the deceleration parameter is measured as $q_{0}=-0.556, -0.554$ for the Model-I and $q_{0}=-0.524, -0.553$ for the Model-II, along two datasets, respectively, which reveals the accelerating stage of the universe expansion. We can obtain the relation $q=0.5(1+3\omega_{de}\Omega_{de})$ for the models with dust fluid ($p=0$), which gives the accelerating phase of the universe for $\omega_{de}\Omega_{de}<-\frac{1}{3}$. Thus, for $\Omega_{de}=0$ (i.e., for $\rho_{de}=0$), $q=0.5>0$, we obtain a decelerating universe, and this confirms that the geometrical modification can explain the accelerating phase of an expanding universe.\\

From Table 3, we can see that for the CC Hubble datasets, the AIC criteria for both models are in the second group, with $2<\Delta IC <6$. This means that our two derived models are in mild tension with the most popular $\Lambda$CDM, while the BIC criteria are in the third group, with $6<\Delta IC <10$ for Model-I and $\Delta IC>10$ for the Model-II. This means that the both models are in mild tension with $\Lambda$CDM \cite{ref114}. Similarly, according to Table 4, for the Pantheon SNe Ia datasets, the AIC criteria suggest that our two derived models are in mild tension with the most favored $\Lambda$CDM, while the BIC criteria depict that the models are strongly disfavored by $\Lambda$CDM.

 %==============================================
\section{Conclusions}
 %==============================================
 
We examine FLRW cosmological models within the framework of Metric-Affine $F(R,Q)$ gravity, as introduced in the publication [arXiv:1205.52666]. In this context, $R$ represents the curvature scalar and $Q$ represents the nonmetricity scalar, both calculated using non-special connections. The updated field equations are derived by employing a flat Friedmann-Lema\^{I}tre-Robertson-Walker (FLRW) metric. In two distinct scenarios involving scalars $u$ and $w$, we establish a correlation between the Hubble constant $H_{0}$, the density parameter $\Omega_{m0}$, and the other model parameters. Subsequently, we employed recently acquired observational datasets, including the cosmic chronometer (CC) Hubble datasets and the Pantheon SNe Ia datasets, to ascertain the most suitable values for the model parameters via MCMC analysis. By utilizing these optimal values of model parameters, we have examined the outcomes and characteristics of the resulting models. Both models we have discovered are transitional phase models and methods for the Lambda CDM model in the late-time universe. We have discovered that the geometric sector's dark equation of state parameter, $\omega_{de}$, exhibits similar behavior to that of a potential dark energy candidate.\\

We considered two specific models, which are known to lead to interesting phenomenology. Our analysis shows that both models are capable of describing the evolution of the universe, supporting observational datasets, namely, cosmic chronometer (CC) Hubble data and Pantheon SNe Ia. We found a fairly large value of $\Omega_{m0}$ and a value of $H_{0}$ that was between the Plank and local estimates but closer to the Plank estimate for both Models I and II, which both use the Lambda CDM paradigm as a particular limit. For the dimensionless parameter $\lambda$, we constrained its value around 0, which shifted towards a positive value due to degeneracy with other parameters $H_{0}$ and $\Omega_{m0}$, etc. For the parameter $\lambda_{0}$ dimensionally equivalent to Hubble constant $H_{0}^{2}$, we constrained its value around the value of the cosmological constant $\Lambda$.\\

We have investigated the behavior of the dark energy EoS parameter $\omega_{de}$ over $z$ with constrained values of model parameters for both models. We observe that for the Model-I, the present values of $\omega_{de}$ fall in the range of phantom and super-phantom regions while for the Model-II, it falls into the quintessential region at late-time. We have also plotted the behavior of $\omega_{de}$ with $z$ in its value. The evolution of $\omega_{de}$ is positive in the early universe for both models, and late-time it converts into negative values that are compatible with the evolution of the deceleration parameter $q(z)$, which depicts the expansion phase of the expanding universe. The Model-I has successfully reached the Lambda CDM stage in the late-time universe while the Model-II shows quintessence scenarios at late-time. Finally, by applying the AIC and BIC criteria, we determined that both Model-I and Model-II were less well-fitted with Lambda CDM cosmological parameters. This is an interesting result since both Model-I and Model-II do not contain $\Lambda$CDM scenarios at present but depict the quintessence, phantom and super-phantom scenarios. Both derived models are transit-phase accelerating universe models which can explain the late-time accelerating scenarios of expanding universe.\\

Since this is the case, we have shown that the $F(R,Q)$ gravity model can explain the accelerated phase of the expanding universe. The derived $F(R,Q)$ gravity model's results are in mild tension with the $\Lambda$CDM standard cosmological model. Furthermore, the $F(R,Q)$ gravity model allows us to recover the original Friedmann model. This $F(R,Q)$ gravity theory is a generalization of both $F(R)$ and $F(Q)$. As a result, the current modified gravity model is intriguing and attracts researchers to reexamine it in order to uncover other cosmological features of this $F(R,Q)$ gravity theory.

%================================================
\section*{Acknowledgments}
%================================================
We are thankful to renowned referees and editors for their valuable suggestions to improve this manuscript. This work was supported by the Ministry of Science and Higher Education of the Republic of Kazakhstan, Grant AP14870191.
%%%%%%%%%%%%%%%%%%%%%%%%%%%%%%%%%%%%%%%%%%%%%%%%%%%%%%%%%%%%
\section{Data Availability Statement}
%%%%%%%%%%%%%%%%%%%%%%%%%%%%%%%%%%%%%%%%%%%%%%%%%%%%%%%%%
No data associated in the manuscript.
%%%%%%%%%%%%%%%%%%%%%%%%%%%%%%%%%%%%%%%%%%%%%%%%%%%%%%%%%%%%%%%%%%%%%%%%%%%%%%%%%%%%%%%%%%%%%%%%%%%%
%%%%%%%%%%%%%%%%%%%%%%%%%%%%%%%%%%%%%%%%%%%%%%%%%%%%%%%%%%%
%%%%%%%%%%%%%%%%%%%%%%%%%%%%%%%%%%%%%%%%%%%%%%%%%%%%%%%%%%%%%%%%%%%%%%%%%%%%%%%%%%%%%%%%%%%%%%%%%%%%%%
\section{Statements and Declarations}
\subsection*{Funding and/or Conflicts of interests/Competing interests}
The author of this article has no conflict of interests. The author have no competing interests to declare that are relevant to the content of this article. Authors have mentioned clearly all received support from the organization for the submitted work.
%%%%%%%%%%%%%%%%%%%%%%%%%%%%%%%%%%%%%%%%%%%%%%%


\begin{thebibliography}{99}
	\bibitem{ref1}	
	C. M. Will, Living Rev. Relativ. \textbf{17}: 4 (2014).
	\bibitem {ref2}
	A. G. Riess, A. V. Filippenko, P. Challis, \textit{et al.}, Astron. J., \textbf{116}: 1009 (1998).
	\bibitem {ref3}
	S. Perlmutter, G. Aldering, G. Goldhaber, \textit{et al.}, Astrophys. J., \textbf{517}: 565 (1999).
	\bibitem {ref4}
	R.A. Knop, G. Aldering, R. Amanullah, \textit{et al.}, Astrophys. J., \textbf{598}: 102 (2003).
	\bibitem {ref5}
	R. Amanullah, C. Lidman, D. Rubin, \textit{et al.}, Astrophys. J., \textbf{716}: 712 (2010).
	\bibitem {ref6}
	D.H. Weinberg, M. J. Mortonsonb, D. J. Eisenstein, \textit{et al.}, Phys. Rep., \textbf{530}: 87 (2013).
	\bibitem {ref7}
	A. Einstein, Naturwissenschaften, \textbf{5}: 770-771 (1917). 
	\bibitem {ref8}
	P. Salucci, N. Turini, and C. Di Paolo, Universe, \textbf{6}: 118 (2020).
	\bibitem {ref9}
	S. Alam \textit{et al.} (BOSS Collaboration), Mon. Not. R. Astron. Soc., \textbf{470}: 2617 (2017). arXiv:1607.03155.
	\bibitem {ref10}
	T.M.C. Abbott \textit{et al.} (DES Collaboration), Phys. Rev. D, \textbf{98}: 043526 (2018).
	\bibitem {ref11}
	M. Tanabashi \textit{et al.} (Particle Data Group), Phys. Rev. D, \textbf{98}: 030001 (2018).
	\bibitem {ref12}
	N. Aghanim \textit{et al.} (Planck Collaboration), Astron. Astrophys., \textbf{641}: A6 (2020).
	\bibitem{ref13}
	E. N. Saridakis, R. Lazkoz, V. Salzano, \textit{et al.},  arXiv:2105.12582v2 [gr-qc].
	\bibitem{ref14}
	T. P. Sotiriou and V. Faraoni, Rev. Mod. Phys., \textbf{82}: 451-497 (2010). arXiv:0805.1726.
	\bibitem{ref15}
	D. Iosifidis, A. C. Petkou, and C. G. Tsagas, Gen. Relativ. Gravit. \textbf{51}: 66 (2019).
	\bibitem{ref16}
	S. Capozziello and S. Vignolo, Annalen der Physik, \textbf{19}: 238-248 (2010).
	\bibitem{ref17}
	R. Aldrovandi and J. G. Pereira, Teleparallel gravity: an introduction, (Springer Science \& Business
	Media, 2012), volume 173, p. 214.
	\bibitem{ref18}
	R. Myrzakulov, Eur. Phys. J. C, \textbf{71}: 1-8 (2011).
	\bibitem {ref19}
	J. Belt\'{r}an Jim\'{e}nez, L. Heisenberg, T. S. Koivisto, \textit{et al.}, Phys. Rev. D, \textbf{2}: 103507 (2020).
	\bibitem{ref20}
	J. M. Nester and H.-J. Yo, arXiv:gr-qc/9809049.
	\bibitem{ref21}
	J. Belt\'{r}an Jim\'{e}nez, L. Heisenberg, and T. S. Koivisto, J. Cosmo. Astropart. Phys., \textbf{2018}: 039 (2018).
	\bibitem {ref22}
	L. Heisenberg, arXiv:2309.15958 [gr-qc].
	\bibitem{ref23}
	N. Bartolo and M. Pietroni, Phys. Rev. D, \textbf{61}: 023518 (1999).
	\bibitem{ref24}
	C. Charmousis, E. J. Copeland, A. Padilla, \textit{et al.}, Phys. Rev. Lett., \textbf{108}: 051101 (2012).
	\bibitem{ref25}
	L. P. Eisenhart, Non-Riemannian geometry, Courier Corporation, (2012).
	\bibitem{ref26}
	F. W. Hehl, J. D. McCrea, E. W. Mielke, \textit{et al.}, Phys. Rep., \textbf{258}: 1-171 (1995).
	\bibitem {ref27}
	T. P. Sotiriou, Class. Quant. Grav., \textbf{26}: 152001 (2009).
	\bibitem {ref28}
	V. Vitagliano, T. P. Sotiriou and S. Liberati, Annals Phys., \textbf{326}: 1259 (2011) [Erratum-ibid. 329, 186 (2013)].
	\bibitem {ref29}
	V. Vitagliano, T. P. Sotiriou and S. Liberati, Phys. Rev. D, \textbf{82}: 084007 (2010).
	\bibitem {ref30}
	F. W. Hehl, E. A. Lord and L. L. Smalley, Gen. Rel. Grav. \textbf{13}: 1037 (1981).
	\bibitem {ref31}
	Vincenzo Vitagliano, Class. Quantum Grav., \textbf{31}: 045006 (2014).
	\bibitem{ref32}
	D. Iosifidis, arXiv:1902.09643.
	\bibitem{ref33}
	D. Iosifidis, Class. Quantum Grav., \textbf{36}: 085001 (2019).	
	\bibitem{ref34}
	D. Iosifidis and T. Koivisto, Universe, \textbf{5}(3): 82 (2019).
	\bibitem{ref35}
	V. Vitagliano, T. P. Sotiriou, and S. Liberati, Annals of Physics, \textbf{326}(5): 1259-1273 (2011).
	\bibitem{ref36}
	T. P. Sotiriou and S. Liberati, Annals of Physics, \textbf{322}(4): 935-966 (2007).
	\bibitem{ref37}
	R. Percacci and E. Sezgin, Phys. Rev. D, \textbf{101}(8): 084040 (2020).
	\bibitem{ref38}
	J. Belt\'{r}an Jim\'{e}nez and A. Delhom, Eur. Phys. J. C, \textbf{80}(6): 585 (2020).
	\bibitem{ref39}
	J. Belt\'{r}an Jim\'{e}nez and A. Delhom, Eur. Phys. J. C, \textbf{79}(8): 656 (2019).
	\bibitem{ref40}
	G. J. Olmo, Int. J. Mod. Phys. D, \textbf{20}: 413-462 (2011).
	\bibitem{ref41}
	K. Aoki and K. Shimada, Phys.	Rev. D, \textbf{100}(4): 044037 (2019).
	\bibitem{ref42}
	F. Cabral, F. S. N. Lobo, and D. Rubiera-Garcia, Universe, \textbf{6}(12): 238 (2020).
	\bibitem{ref43}
	S. Ariwahjoedi, A. Suroso, and F. P. Zen, Class. Quantum Grav., \textbf{38}: 155009 (2021).
	\bibitem{ref44}
	J.-Z. Yang, S. Shahidi, T. Harko, \textit{et al.}, Eur. Phys. J. C, \textbf{81}(2): 111 (2021).
	\bibitem{ref45}
	T. Helpin and M. S. Volkov, Int. J. Mod. Phys. A, \textbf{35}(02n03): 2040010 (2020).
	\bibitem{ref46}
	S. Bahamonde and J. G. Valcarcel, J. Cosmo. Astropart. Phys., \textbf{2020}(09): 057 (2020).
	\bibitem{ref47}
	D. Iosifidis and L. Ravera, Class. Quantum Grav., \textbf{38}(11): 115003 (2021).
	\bibitem{ref48}
	D. Iosifidis, Class. Quantum Grav., \textbf{38}: 195028 (2021). arXiv:2104.10192.
	\bibitem{ref49}
	D. Iosifidis, Class. Quantum Grav., \textbf{38}(1): 015015 (2020).
	\bibitem{ref50}
	D. Iosifidis, Eur. Phys. J. C, \textbf{80}(11): 1042 (2020).
	\bibitem{ref51}
	D. Iosifidis and L. Ravera, Eur. Phys. J. C, \textbf{81}: 736 (2021).
	\bibitem{ref52}
	J. Belt\'{r}an Jim\'{e}nez and T. S. Koivisto, Phys. Lett. B, \textbf{756}: 400–404 (2016).
	\bibitem{ref53}
	J. Belt\'{r}an Jim\'{e}nez and T. S. Koivisto, Universe, \textbf{3}(2): 47 (2017).
	\bibitem{ref54}
	D. Kranas, C. G. Tsagas, J. D. Barrow, \textit{et al.}, Eur. Phys. J. C, \textbf{79}(4): 341 (2019).
	\bibitem{ref55}
	C. Barrag\'{a}n, G. J. Olmo, and H. Sanchis-Alepuz, Phys. Rev. D, \textbf{80}(2): 024016 (2009).
	\bibitem{ref56}
	K. Shimada, K. Aoki, and Kei-ichi Maeda, Phys. Rev. D, \textbf{99}(10): 104020 (2019).
	\bibitem{ref57}
	M. Kubota, Kin-ya Oda, K. Shimada, \textit{et al.}, J. Cosmo. Astropart. Phys., \textbf{2021}(03): 006 (2021).
	\bibitem{ref58}
	Y. Mikura, Y. Tada, and S. Yokoyama, EPL, \textbf{132}(3): 39001 (2020).
	\bibitem{ref59}
	Y. Mikura, Y. Tada, and S. Yokoyama, Phys. Rev. D, \textbf{103}: 101303 (2021). arXiv:2103.13045.
	\bibitem{ref60}
	F. W. Hehl, G. D. Kerlick, and P. von der Heyde, Zeitschrift fuer Naturforschung A, \textbf{31}(2): 111-114 (1976).
	\bibitem{ref61}
	OV Babourova and BN Frolov, arXiv:gr-qc/9509013.
	\bibitem{ref62}
	Y. N. Obukhov and R. Tresguerres, Phys. Lett. A, \textbf{184}(1): 17-22 (1993).
	\bibitem{ref63}
	D. Iosifidis, JCAP, \textbf{04}: 072 (2021).
	\bibitem {ref64}
	A. Conroy and T. Koivisto, Eur. Phys. J. C, \textbf{78}: 923 (2018). arXiv:1710.05708.
	\bibitem {ref65}
	R. Myrzakulov, Eur. Phys. J. C, \textbf{72}: 2203 (2012). arXiv:1207.1039.
	\bibitem {ref66}
	E. N. Saridakis, S. Myrzakul, K. Myrzakulov, \textit{et al.}, Phys. Rev. D, \textbf{102}: 023525 (2020). arXiv:1912.03882.
	\bibitem {ref67}
	M. Jamil, D. Momeni, M. Raza, \textit{et al.}, Eur. Phys. J. C, \textbf{72}: 1999 (2012). arXiv:1107.5807.
	\bibitem {ref68}
	M. Sharif, S. Rani and R. Myrzakulov, Eur. Phys. J. Plus, \textbf{128}: 123 (2013). arXiv:1210.2714.
	\bibitem {ref69}
	S. Capozziello, M. De Laurentis and R. Myrzakulov, Int. J. Geom. Meth. Mod. Phys., \textbf{12}: 1550095 (2015) [arXiv:1412.1471].
	\bibitem {ref70}
	P. Feola, X. J. Forteza, S. Capozziello, \textit{et al.}, arXiv:1909.08847.		
	\bibitem {ref71}
	F.K. Anagnostopoulos, S. Basilakos, and E.N. Saridakis, Phys. Rev. D, \textbf{103}: 104013 (2021). arXiv:2012.06524[gr-qc].
	\bibitem {ref72}
	N. Myrzakulov, R. Myrzakulov, L. Ravera, arXiv:2108.00957.
	\bibitem {ref73}
	D. Iosifidis, N. Myrzakulov, R. Myrzakulov, Universe, \textbf{7}: 262 (2021). arXiv:2106.05083.
	\bibitem {ref74}
	T. Harko, N. Myrzakulov, R. Myrzakulov, \textit{et al.}, arxiv:2110.00358v1.
	\bibitem {ref75}
	R. Saleem, Aqsa Saleem, Chin. J. Phys., \textbf{84}: 471-485 (2023).
	\bibitem {ref76}
	D. Iosifidis, R. Myrzakulov, L. Ravera, \textit{et al.}, arXiv:2111.14214.
	\bibitem {ref77}
	G. Papagiannopoulos, S. Basilakos, E.N. Saridakis, arXiv:2202.10871.
	\bibitem {ref78}
	S. Kazempour, A. R. Akbarieh, arXiv:2309.09230.
	\bibitem{ref79} 
	D. C. Maurya, R. Myrzakulov, Eur. Phys. J. C, \textbf{84}: 534 (2024). arXiv:2401.00686.
	\bibitem{ref80} 
	D. C. Maurya, R. Myrzakulov, Eur. Phys. J. C, \textbf{84}: 625 (2024). arXiv:2402.02123.
	\bibitem{ref81}
	D. C. Maurya, K. Yesmakhanova, R. Myrzakulov, \textit{et al.}, arXiv:2404.09698[gr-qc].
	\bibitem {ref82}
	S. Capozziello, V. De Falco, C. Ferrara, Eur. Phys. J. C, \textbf{82}: 865 (2022).
	\bibitem {ref83}
	K. Shimada, K. Aoki, Kei-ichi Maeda, Phys. Rev. D, \textbf{99}: 104020 (2019).
	\bibitem {ref84}
	D. A. Gomes, J. Beltr{\'a}n Jim{\'e}nez, A. Jim{\'e}nez Cano, \textit{et al.}, Phys. Rev. Lett., \textbf{132}: 141401 (2024).	
    \bibitem{ref85} 
    R. Myrzakulov, arXiv:1205.5266v6[physics.gen-ph].        
    \bibitem{ref86} 
    K. Yesmakhanova, N. Myrzakulov, S. Myrzakul, \textit{et al.}, arXiv:2101.05318.
    \bibitem {ref87}
    D. Foreman-Mackey, D.W. Hogg, D. Lang, \textit{et al.}, Publ. Astron. Soc. Pac., \textbf{125}: 306 (2013). https://doi.org/10.1086/670067
    \bibitem {ref88}
    J. Simon, L. Verde, R. Jimenez, Phys. Rev. D, \textbf{71}: 123001 (2005). https://doi.org/10.1103/PhysRevD.71.123001
    \bibitem {ref89}
    G.S. Sharov, V.O. Vasiliev, Math. Model. Geom., \textbf{6}: 1-20 (2018). https://doi.org/10.26456/mmg/2018-611
    \bibitem {ref90}
    M. Moresco, R. Jimenez, L. Verde, \textit{et al.}, ApJ, \textbf{898}: 82 (2020).
    \bibitem {ref91}
    G. Ellis, R. Maartens, M. MacCallum, Relativistic Cosmology (Cambridge University Press, Cambridge, 2012). https://doi.org/10.1017/CBO9781139014403.
    \bibitem {ref92}
    K. Asvesta, L. Kazantzidis, L. Perivolaropoulos, \textit{et al.}, Mon. Not. R. Astron. Soc., \textbf{513}: 2394-2406 (2022). https://doi.org/10.1093/mnras/stac922.
    \bibitem {ref93}
    D. M. Scolnic, D. O. Jones, A. Rest, \textit{et al.}, Astrophys. J., \textbf{859}: (2018) 101.
    \bibitem {ref94}
    R. Andrae, T. Schulze-Hartung and P. Melchior, arXiv:1012.3754.
    \bibitem {ref95}
    K. Anderson, Model selection and multimodel inference: a practical information-theoretic approach, Second Edition (Springer, New York (2002)).
    \bibitem {ref96}
    K. P. Burnham, D. R. Anderson, Sociological Methods \& Research, \textbf{33}(2): 261-304 (2004).
    \bibitem {ref97}
    A. R. Liddle, Mon. Not. Roy. Astron. Soc., \textbf{377}: L74 (2007). arXiv:astro-ph/0701113.
    \bibitem {ref98}
    R. E. Kass and A. E. Raftery, J. Am. Statist. Assoc., \textbf{90}(430): 773 (1995).
    \bibitem {ref99}
    S. Cao and B. Ratra,Phys. Rev. D, \textbf{107}: 103521 (2023). arXiv:2302.14203[astro-ph.CO].		
    \bibitem {ref100}
    S. Cao and B. Ratra, MNRAS, \textbf{513}: 5686-5700 (2022). arXiv:2203.10825[astro-ph.CO].
    \bibitem {ref101}
    A. Dom\'{\i}nguez, R. Wojtak, J. Finke, \textit{et al.}, ApJ, \textbf{885}: 137 (2019). arXiv:1903.12097v2[astro-ph.CO].
    \bibitem {ref102}
    Chan-Gyung Park, Bharat Ratra, Phys. Rev. D, \textbf{101}: 083508 (2020). arXiv:1908.08477[astro-ph.CO].
    \bibitem {ref103}
    W. Lin and M. Ishak, JCAP, \textbf{2105}:009 (2021). arXiv:1909.10991v3[astro-ph.CO].
    \bibitem {ref104}
    W. L. Freedman,  B. F. Madore, T. Hoyt, \textit{et al.}, ApJ, \textbf{891}: 57 (2020). arXiv:2002.01550v1[astro-ph.GA].
    \bibitem {ref105}
    S. S. Boruah, M. J. Hudson, G. Lavaux, MNRAS, \textbf{507}2: 2697-2713 (2021). arXiv:2010.01119v1[astro-ph.CO].
    \bibitem {ref106}
    Wendy L. Freedman, ApJ, \textbf{919}: 16 (2021). arXiv:2106.15656v1[astro-ph.CO].
    \bibitem {ref107}
    Q. Wu, G. Q. Zhang, F. Y. Wang, MNRAS, \textbf{515} (1): L1-L5 (2022). arXiv:2108.00581v2[astro-ph.CO].
    \bibitem{ref108}
    N. Aghanim, \textit{et al} (Planck Collaboration), A \& A, \textbf{641}: A6 (2020). arXiv:1807.06209[astroph.CO].
    \bibitem{ref109}
    A. G. Riess, S. Casertano, W. Yuan, \textit{et al.}, ApJ, \textbf{908}(1): L6 (2021). arXiv:2012.08534.
    \bibitem{ref110}
    A. Pradhan, D. C. Maurya, G. K. Goswami, \textit{et al.}, Int. J. Geom. Meth. Mod. Phys., \textbf{20}: 2350105 (2023).
    \bibitem{ref111}
    D. C. Maurya, Phys. Dark Univ., \textbf{42}: 101373 (2023). https://doi.org/10.1016/j.dark.2023.101373.
    \bibitem {ref112}
    D. Brout, D. Scolnic, B. Popovic, \textit{et al.}, ApJ, \textbf{938}: 110 (2022).
    \bibitem {ref113}
    A. R. Lalke, G. P. Singh and A. Singh, Eur. Phys. J. Plus, \textbf{139}:288 (2024). https://doi.org/10.1140/epjp/s13360-024-05091-5.
    \bibitem {ref114}
    B. Efron, A. Gous, R. E. Kass, \textit{et al.}, Model Selection IMS Lecture Notes - Monograph Series, \textbf{38}: 208-256 (2001).
\end{thebibliography}
 \end{document}